\documentclass[11pt,a4paper]{article}
\pdfoutput=1
\usepackage{jheppub}

\usepackage{amsmath}
\usepackage{verbatim}
\usepackage{graphicx}
\usepackage{mathrsfs}
\usepackage{appendix}
\usepackage{caption}
\usepackage{float}
\usepackage{subfigure}
\usepackage{epstopdf}
\usepackage{multicol}
\usepackage{tikz}

\newcommand{\e}{\epsilon}

\newcommand{\be}[1]{ \begin{equation}\label{#1} }
	\newcommand{\ee}{\end{equation}}
\newcommand{\bea}[1]{\begin{eqnarray}\label{#1} }
	\newcommand{\eea}{\end{eqnarray}}
\newcommand{\bes}{\begin{subequations}}
	\newcommand{\ees}{\end{subequations}}

\newcommand{\p}{\partial}

\renewcommand{\b}{\beta}

\newcommand{\D}{\Delta}

\newcommand{\non}{\nonumber}

\newcommand{\ie}{\emph{i.e.}}

\title{Carrollian Yang-Mills Theory}

\author{Minhajul Islam} \author{\\}

\affiliation{Indian Institute of Technology Kanpur, Kalyanpur, Kanpur 208016. INDIA. \\}

\emailAdd{ minhajul@iitk.ac.in}

\abstract{By doing a small $c$ (speed of light) expansion of $SU(N)$ Yang-Mills fields, we construct two different electric and two different magnetic sectors actions of Carrollian Yang-Mills theory. For both electric and magnetic cases, one sector contains non-trivial self-interaction, and another is $N^2-1$ copies of respective sector Carrollian abelian theory. In $d=4$ , all the four sectors are invariant under infinite Carrollian Conformal symmetry. There are no central extensions when analyzing charge algebra at the phase space level. Lastly, we compute propagators for all four sectors and vertices for two non-trivial sectors. Propagators in position space show ultra-local behavior.
}

\begin{document}
	
	\maketitle
	\section{Introduction}
	
	The construction of the spectacularly successful Standard Model of particle physics, which describes nature around us, is based on the foundation of relativistic quantum field theory (QFT). But, often, to describe real life systems, it is desirable to look at approximations and limits of the more fundamental theory. 
	
	\medskip
	
	Gauge theories are the backbone of theoretical physics. Three of the four fundamental forces of nature are explained by Yang-Mills theory. Even the first example of the most promising formalism to understand Quantum gravity, called AdS/CFT holographic duality, is constructed using a supersymmetric version of Yang-Mills theory \cite{Maldacena:1997re} . The AdS/CFT holographic duality relates the $d + 1$-dimensional gravitational theory to the d-dimensional field theory. More specifically, \cite{Maldacena:1997re} connects a string theory living on five-dimensional Anti-de Sitter (AdS) spacetimes (times a five-sphere) and ${\cal N} = 4$ $SU(N)$ Supersymmetric Yang-Mills (SYM) theory which is a four-dimensional conformal field theory living on the boundary of AdS. 
	
	\medskip
	
	In this paper, we will look at Yang-Mills theories from a different perspective. We will attempt to understand the theory in the limit when the speed of light goes to zero. The diametrically opposite limit, where $c\rightarrow \infty$ is clearly of physical interest as it describes Galilean or non-relativistic (NR) physics, and is useful to describe a range of day to day physical systems like hydrodynamics. Below we clarify why the other limit, called the Carrollian limit, is important. 
	
	
	\medskip
	
	If we adopt a group-theoretic approach to understand QFT at these two different (Galilean and Carrollian) limits, we would begin from the Poincar\'e algebra and take the large c (speed of light) limit and small c limit. The two symmetry algebras that would be obtained as a result are different and are the familiar Galilean algebra, and the not-so-familiar Carrollian algebra. In both these limits, many interesting counter-intuitive concepts emerge. In both cases, spacetime metrics degenerate, light-cones open up for non-relativistic theory and close up for Carrollian theory, and symmetry algebra gets enhanced. 
	
	\medskip
	
	Non-relativistic theories, corresponding to $c\rightarrow \infty$, are important for condensed matter physics, non-AdS holography, and hydrodynamics. In this limit, as mentioned previously the metric degenerates, spacetime loses its Reimmanian structure, and a new spacetime structure emerges called Newton-Cartan spacetime. Selected references on the construction of  non-relativistic field theories and related Newton-Cartan spacetime structures are \cite{Duval:1984cj,Duval:2009vt,Bleeken:2015ykr,Bergshoeff:2015sic,Hansen:2020pqs}. In connection with the construction of symmetries, one of the interesting techniques to construct non-relativistic physics is to start from a Poincar\'e invariant theory and do a large c-expansion. Using this approach we get many interesting insights into non-relativistic physics like order-wise enhanced symmetry algebra, and actions \cite{Hansen:2019svu,Hansen:2020wqw, Hansen:2020pqs, Ergen:2020yop}.
	
	\medskip    
	
	
	
	Our main focus in this paper is the other limit corresponding to $c\rightarrow 0$, which is called the Carrollian limit.  At first sight, sending the speed of light to zero may seem unnatural and the expectation is that this would lead to unphysical models. But recently, this particular limit has been resurgent with different applications, mainly connected to the understanding of flat space holography \cite{Susskind:1998vk}. As mentioned before, one of the most promising tools to understand Quantum gravity is the AdS/CFT duality. In the limit of infinite radius of curvature, AdS spacetime become flat spacetime. On the dual side, the infinite radius limit corresponds to sending the speed of light to zero \cite{Bagchi:2012cy}. The boundary theory thus becomes a Carrollian conformal field theory. Some important references for holography for asymptotically flat spacetime are \cite{Susskind:1998vk,Bagchi:2010eg,Bagchi:2012cy, Bagchi:2014iea,Bagchi:2013qva,Bagchi:2016bcd,Barnich:2006av,Barnich:2010eb,Barnich:2012aw,Barnich:2012xq,Ciambelli:2018wre,Bagchi:2022nvj}. The understanding of flat space holography recently has taken two different directions, viz. Celestial holography and Carrollian holography. Celestial holography relates gravity in 4d asymptotically flat spacetimes to a 2d CFT living on the celestial sphere \cite{Pasterski:2021raf,Raclariu:2021zjz,Pasterski:2021rjz}. On the other hand, Carrollian holography relates 4d asymptotically flat gravity to 3d Carrollian CFTs living on the entire null boundary of 4d bulk spacetime \cite{Dappiaggi:2004kv,Dappiaggi:2005ci,Bagchi:2016bcd,Bagchi:2019xfx,Bagchi:2019clu,Bagchi:2022owq,Duval:2014uva,Duval:2014lpa}. Recently, some fascinating works have been done to connect both formalisms \cite{Bagchi:2022emh,Donnay:2022aba}.

	The most successful example of AdS/CFT is the original Maldacena correspondence relating $\mathcal{N}=4$ $SU(N)$  Supersymmetric Yang-Mills theory in $d=4$ to gravity in AdS$_5$. One of our long-term goals is to understand the flatspace version of the Maldacena correspondence. As an important intermediate step, we wish to construct the Carrollian version of Super-Yang-Mills theory. This is the main motivation for constructing Carrollian Yang-Mills (CYM) theory and, in particular, actions for CYM  in this paper. 
	\medskip
	
	Carrollian physics has also emerged in other interesting places and here we quickly summarize these exciting developments. Carrollian structure appear on any null hyper-surface. Every black hole solutions of general relativity contains a horizon that is nothing but a null surface. Carrollian structures on black hole horizons have been considered in \cite{Donnay:2019jiz}. Carrollian gravity may provide a tractable limit of general relativity and be useful for various physical context. This has been studied in \cite{Dautcourt:1997hb,Hartong:2015xda,Bergshoeff:2017btm,Duval:2014uoa,Ciambelli:2018ojf,Morand:2018tke,Ciambelli:2019lap}. Carroll theory is also important for cosmology, inflation \cite{deBoer:2021jej}, fluid mechanics \cite{Ciambelli:2018ojf,Ciambelli:2018xat,Ciambelli:2018wre,Ciambelli:2019lap,Petkou:2022bmz, Freidel:2022bai,Freidel:2022vjq,Redondo-Yuste:2022czg}, fractons \cite{Nandkishore:2018sel,Bidussi:2021nmp,Perez:2022kax}, flat physics in condensed matter systems \cite{Bagchi:2022eui}. Inspired by large $c$-expansion and construction of non-relativistic physics, small $c$-expansion was introduced to understand Carrollian physics in \cite{Hansen:2021fxi}. Finally, the Carrollian limit of the string theory worldsheet leads to the very high energy tensionless regime of strings. This has been investigated in detail in \cite{Bagchi:2013bga,Bagchi:2015nca,Bagchi:2020fpr,Bagchi:2021rfw,Bagchi:2021ban}.  Recently there has been some interesting work done on Carroll fermions \cite{Bagchi:2022eui,Banerjee:2022ocj,Yu:2022bcp,Hao:2022xhq}. 
	
	\medskip
	
	Before moving on to Carrollian gauge theories, which will be the focus in this paper, we briefly recall previous works on Galilean gauge theories. Galilean gauge theory for $U(1)$ theory was first constructed long ago \cite{LBLL}. In \cite{Bagchi:2014ysa,Bagchi:2015qcw,Bagchi:2017yvj} authors realized infinite-dimensional Galilean conformal symmetry at the level of equations of motion in Galilean abelian and Galilean Yang-Mills theory. Subsequently there is some detailed work on action constructions for both Galilean abelian \cite{Festuccia:2016caf,Banerjee:2019axy} and Yang-Mills theory \cite{Bagchi:2022twx}. Quantum properties of Galilean scalar electrodynamics were studied in \cite{Chapman:2020vtn} and that of Galilean QED in \cite{Banerjee:2022uqj}.  
	
	\medskip
	
	The Carrollian algebra was first discussed in \cite{LevyLeblond,NDS}. More recently Carroll conformal structures have been analyzed at the level equations of motion in \cite{Bagchi:2016bcd,Bagchi:2019clu,Bagchi:2019xfx,Bagchi:2022owq}. In \cite{Basu:2018dub} Carrollian action was constructed for the so-called electric abelian theory, which is an interacting field theory with scalar field \cite{Banerjee:2020qjj,deBoer:2021jej}. Using the small $c$-expansion, the magnetic sector of Carrollian abelian theory has been recently constructed \cite{deBoer:2021jej}. The conformal structure of this magnetic action was analyzed. In \cite{Henneaux:2021yzg} authors constructed off-shell Carrollian Yang-Mills theory in the Hamiltonian formulation. However, at present there is no action formulation for the Carrollian Yang-Mills theory. 
	
	\medskip
	
	In this paper, we construct Carrollian Yang-Mills actions using the small c-expansion technique. We find four different sectors of Carrollian Yang-Mills theory. This construction depends on the power of c we consider during field expansion. All four sectors exhibit infinite Carrollian conformal invariance in four spacetime dimensions. The energy-momentum tensors for all four sectors are analyzed, and their conservation is established using equations of motion and Bianchi identities. To see charge algebra, we calculate charges for all the four sectors and show that the symmetry is realized at the level of charge algebra. We begin our investigation of the quantum properties of the theory and calculate all the propagators and vertices.
	A detailed quantum mechanical analysis is kept for future work. 
	
	\medskip

	\section*{Outline of the paper}
	The paper is organized as follows. We begin in Sec.\ref{CCA and rep} with a review of Carrollian conformal algebra (CCA). After that, we talk about an infinite extension of the CCA. 
	
	In Sec.\ref{YMACE} we address relativistic Yang-Mills theory and its small $c$-expansion. We take expansion of fields as $A_{\mu}^{a}=\sum_{n=0}^{\infty} c^{\lambda}c^{2n}A_{\mu}^{a(n)}$, where $\lambda$ is a non-negative constant parameter.  Using $\lambda=0$, we get the electric and the magnetic sectors of CYM with a non-trivial term or self-interaction term. For $\lambda$ with any non-zero value, we get copies of the abelian electric and the abelian magnetic sectors. Here for any non-zero value, we choose the lowest even integer value two, which is explained in detail.
	
	In Sec.\ref{CYM all actions}, we address details of all the sectors of CYM action.
	For each sector, firstly, we give the action in a compact form, and write the equations of motion, and the gauge symmetry. After that, we show its invariance under infinite CCA in four spacetime dimensions. Finally, we analyze the energy-momentum tensor with its improved version and its conversation. 
	In Sec.\ref{Noether charge and algebra}, we calculate Noether charges and check the charge algebra for these actions. 
	
	In Sec.\ref{Quantum} we briefly discuss Feynman rules for propagators and vertices for all the four sectors along with the Feynman diagrams. In this section, we also talk about propagators in position space. In Sec.\ref{CF} we conclude with a summary of our results and a list of future directions.

	\section{Carrollian Conformal Algebra and Representation}\label{CCA and rep}
	The UR or Carrollian symmetry can be obtained by  performing  an In\"{o}n\"{u}--Wigner
	contraction on the relativistic conformal generators. The corresponding contraction of the spacetime coordinates for a $d$-dimensional CFT is described as
	\be{stscale}
	x_i \to x_i, \qquad t \to \e\, t; \qquad \e \to 0\,.
	\ee
	Here, $i$ runs over the spatial coordinates $i=1,\hdots,d-1$. The above contraction can also be interpreted as taking the limit of vanishing speed of light, $c\to 0$. The Carrollian generators are obtained by performing the space-time contraction on the parent relativistic generators. For example, we obtain Carrollian boost generator 
	\be{convensca}
	B_i^{\textrm{\tiny rel}}=  -x_i \p_t-t\p_i \xrightarrow[]{\text{using \eqref{stscale}}} -\frac{1}{\e}x_i \p_t-t\p_i 
	\xrightarrow[]{\text{redefined}\, B_i}  B_i= 
	\displaystyle \lim_{\e \rightarrow 0}\e B_i^{\textrm{\tiny rel}} 
	\xrightarrow[\text{limit}]{\text{Carroll}} B_i=- x_i \p_t\,.
	\ee
	The other Carrollian generators are also obtained by doing the analysis like above. They are given by
	\bes\label{genearl}
	\begin{eqnarray}
		&& H = \p_t, \quad  B_i=-x_i \p_t, \quad  K_i = -2 x_j (t\p_t+x_i\p_i)+x_j x_j \p_i, \quad K =x_i x_i \p_t, \\
		&& D=-(t\p_t+x_i \p_i), \quad P_i=\p_i, \quad J_{ij}=-(x_i\p_j-x_j\p_i)\,. 
	\end{eqnarray}
	\ees
	These generate the finite Conformal Carrollian Algebra (f-CCA), which is $iso(d,1)$ for a $d$-dimensional field theory \cite{Bagchi:2016bcd,Bagchi:2019xfx}:
		\begin{align}\label{algebra}
			[J_{ij}, B_k ]&=\delta_{k[i}B_{j]}, & [J_{ij}, P_k ]&=\delta_{k[i}P_{j]}, & [J_{ij}, K_k ]&=\delta_{k[i}K_{j]}, & [B_i,P_j]&=\delta_{ij}H, \nonumber \\
			[B_i,K_j]&=\delta_{ij}K, & [D,K]&=-K, & [K,P_i]&=2B_i, & [K_i,P_j]&=-2D\delta_{ij}-2J_{ij}, \nonumber \\ [H,K_i]&=2B_i, & [D,H]&=H, & [D,P_i]&=P_i, & [D,K_i]&=-K_i.
		\end{align}
	The sub-algebra consisting of the generators $\{J_{ij}, B_i, P_i, H\}$ forms the $c\to0$ limit of the Poincar{\'e} algebra {\it viz.} the Carrollian algebra \cite{Leblond65}. 
	
	\medskip
	
	\noindent
	Unlike the relativistic conformal algebra, even in dimensions greater than two, it is possible to give the finite algebra in \eqref{algebra} an infinite-dimensional lift by introducing time translation generator with arbitrary spatial dependence
	\be{2}
	M_f=f(x_i)\p_t\,.
	\ee
	Here, $M_f$ generates the infinite set of super-translations. In the above expression  $f(x_i)$ is an arbitrary function of the spatial co-ordinates $x_i$, which we restrict to polynomials. We obtain the finite generators of f-CCA, i.e., $M_f = H,B_i,K$ when $f(x_i)=1,-x_i,x_k x_k$ respectively.  The super-translation generators $M_f$ along with the finite set of generators $\{B_i,J_{ij},H,P_i,D,K,K_i\}$  describe the infinite-dimensional CCA. For $d\geq 4$ it can be written as \cite{Basu:2018dub,Bagchi:2016bcd}:
	\begin{subequations}
		\label{infinitealgebra1}
		\begin{align}
			[P_i, M_f] &=M_{\p_i f}, & \quad [D,M_f]  &=M_{(-x_i \p_i f+f)},\\
			[K_i,M_f]&= M_{2x_i f+x_k x_k\p_i f-2x_ix_k\p_k f}, &\quad [J_{ij},M_f]&= M_{-x_{[i}\p_{j]}f}\,.
		\end{align}
	\end{subequations}
	For more details of the algebraic aspects of Carrollian conformal symmetry, the reader is pointed to \cite{Bagchi:2016bcd}. In this paper our focus is on spacetime dimension $d=4$.

	\subsection*{Representation theory}\label{repcarroll} 
	The representation theory of the CCA based on highest weights was first constructed in \cite{Bagchi:2016bcd}. Further analysis on representation extended to fields of different integer and half-integer spins was given in \cite{Bagchi:2019xfx}. For the CCA, the states are labeled with the eigenvalues of rotation and dilatation generators. The  construction of representation is summarized below.
	
	\medskip
	
	\noindent 
	The Carrollian CFT fields are labeled with scaling dimension $\Delta$ and spin $j$ as
	\be{4}
	[D,\Phi(0,0)]=\Delta \Phi(0,0),\qquad\qquad [J^2, \Phi(0,0)]=j(j+1)\Phi(0,0)\,.
	\ee 
	The action on a generic field of Carrollian rotation, space- and time-translation is given by
	\be{5}
	[J_{ij},\Phi(0,0)]=\Sigma_{ij}\Phi(0,0),\quad[H,\Phi(t,x_i)]=\p_t \Phi(t,x_i),\quad[P_i,\Phi(t,x_i)]=\p_i \Phi(t,x_i)\,.
	\ee
	The Carrollian conformal primaries are defined as 
	\be{6}
	[K_i,\Phi(0,0)]=0,\;\; [K,\Phi(0,0)]=0,~~[M_{f},\Phi(0,0)]=0~~\text{for polynomial degree} > 1\,.
	\ee
	The Carrollian boost acts on the primary non-trivially because the fields are not eigenstates of Carrollian boosts. The transformation of a generic field under Carrollian boosts can be written using the Jacobi identity. The action of Carroll boost on the fields is
	\be{7}
	[B_i,\Phi(0,0)]=r\varphi_i+\, s \sigma_i\phi + s^{\prime} \sigma_i\chi\, + a A_t \delta_{ji}+b A_i+ \hdots,
	\ee
	where $\varphi, \,\{\phi,\chi \},\,  \{A_t , A_k\}$ denote the primary fields of different spins $(0,\frac{1}{2},1)$. The constants $r,\{s,s^\prime\},\{a,b\}$ cannot be determined just from the symmetries, but can only be fixed though dynamics. One way to determine them   is the limit $c\to0$ of the dynamics of the corresponding relativistic theory. The above action of the Carroll boost can be generalized for any spin theory.
	
	\medskip
	
	\noindent 
	We use the conventional way to define a primary field $\Phi(t,x_i)$  for the CCA at any spacetime point from the origin as
	\be{8}
	\Phi(t,x)=U \Phi(0,0) U^{-1}, \quad \text{where} ~ U=e^{-tH-x_i P_i}.
	\ee
	The action of all the generators of the finite and infinite CCA on this generic Carrollian primary $\Phi(t,x_i)$ can be written as
	\begin{subequations}
		\label{repgen}
		\begin{align}
			[J_{ij}, \Phi(t,x_i)]
			&= (x_i \p_j-x_j \p_i ) \Phi(t,x)+\Sigma_{ij}\Phi(t,x_i),\\
			[B_j, \Phi(t,x_i)]
			&=x_j\p_t \Phi(t,x)-U [B_j, \Phi(0,0)]U^{-1},\\ 
			[D, \Phi(t,x_i) ]
			&= (t\p_t+x_i \p_i+\D) \Phi(t,x_i),\\ 
			[K_j, \Phi(t,x_i) ]
			&=  (2\Delta x_j+2x_jt\partial_t+2x_i x_j \partial_i-2x_i \Sigma_{ij}- x_i x_i \partial_j )\,\Phi(t,x)\non\\
			&\quad -2t \,U [B_j, \Phi(0,0)]U^{-1},\\ 
			[M_f, \Phi(t,x) ] &= f(x_i)\p_t \Phi(t,x)+\p_j f\:U [B_j, \Phi(0,0)]U^{-1}.
		\end{align}
	\end{subequations}
	This is a summary of CCA and its representation, which we have used extensively in the following sections.  We will see for our example what are the constants used in equation \eqref{7}.
	\section{Yang-Mills action and small c-expansion}\label{YMACE}
	
	
	The Yang-Mills theory in $(d+1)$-dimensions is described by the action
	\begin{eqnarray}\label{yma}
		\mathcal{S}_{YM} = \int d^{d+1}x	\,\mathcal{L}_{YM}= \int d^{d+1}x\,\Big(-\frac{1}{4}F^{{\mu} {\nu} a}F_{{\mu} {\nu}}^{ a} \Big),
	\end{eqnarray}
	and the equations of motion
	\begin{eqnarray}\label{eomym}
		\partial_{{\mu}}F^{{\mu}{\nu} a}+gf^{abc}A_{{\mu}}^{b}F^{{\mu}\tilde{\nu} c}=0,
	\end{eqnarray}
	where $a=1,2,...,N^2 -1$. The non-abelian field strength ($F_{\mu\nu}^{a}$) is defined as $F_{{\mu} {\nu}}^a=\p_{{\mu}}A_{{\nu}}^a-\p_{{\nu}}A_{{\mu}}^a+gf^{abc}A_{{\mu}}^bA_{{\nu}}^c$. Here, $A^{a}_{{\mu}}$ is the gauge field and $f^{abc}$ are structure constants of the underlying gauge group.
	
	If we write the above action making the speed of light $c$ explicit, the resulting action is \footnote{$x^{0}=ct$ so $\p_{0}=\frac{1}{c}\p_{t}$ and $A_{0}=\frac{1}{c}A_{t}$}
	\begin{eqnarray}
		S=\int d^{d+1}x\bigg(+\frac{1}{2c^{2}}F_{ti}^{ a}F_{ti}^{a}-\frac{1}{4}F^{ij a}F_{ij}^{a}\bigg).
	\end{eqnarray}
	The action is divided into two parts. The first part contains the temporal component of the gauge field $(A_{t}^{a})$ along with the spatial components $(A_{i}^{a})$. The second part is just dependent on the spatial components $(A_{i}^{a})$ of the gauge field.
	To proceed with the small $c$-expansion, we write the gauge fields as an expansion in $c$ as
	\begin{eqnarray}
		&& A_{t}^{a}=
		\sum_{n=0}^{\infty}c^{\lambda}c^{2n}a_{t}^{a(n)}, \quad
		A_{i}^{a}=
		\sum_{n=0}^{\infty}c^{\lambda}c^{2n}a_{i}^{a(n)}. \label{carroll expansion of fields}
	\end{eqnarray}
	Using these expansions, the first part of the Lagrangian is
	\begin{eqnarray} \label{time part carroll expansion}
		&& \hspace{-8mm} \frac{1}{2c^{2}}F_{ti}^{ a}F_{ti}^{a} = \frac{1}{2}\bigg[c^{2\lambda-2}\sum_{n,m=0}^{\infty}(\p_{t}a_{i}^{a(n)}-\p_{i}a_{t}^{a(n)})(\p_{t}a_{i}^{a(m)}-\p_{i}a_{t}^{a(m)}) +c^{3\lambda-2}\sum_{n,m,l=0}^{\infty}2gf^{abc} \nonumber \\
		&& \hspace{10mm} (\p_{t}a_{i}^{a{(n)}} -\p_{i}a_{t}^{a(n)})a_{t}^{b(m)}a_{i}^{c(l)}  +c^{4\lambda-2}\sum_{n,m,l,p=0}^{\infty}g^{2}f^{abc}f^{ade}a_{t}^{b(n)}a_{i}^{c(m)}a_{t}^{d(l)}a_{i}^{e(p)}\bigg].
	\end{eqnarray}
If we look at only the first term above, \ie\ for the abelian case, $c^{\lambda}$ becomes an overall factor and there will be only one result for different $\lambda$ \cite{deBoer:2021jej}. However, because of the self-interaction terms of the gauge fields in the second and third terms, we can not take out $c^{\lambda}$ as an overall factor. This leads to distinct sectors of Carroll invariant non-abelian gauge theories corresponding to  $\lambda = 0$ and $\lambda\neq 0$.
Similarly, the second part (fully spatial part) of the Lagrangian is
	\begin{eqnarray}\label{Carroll expansion space part}
		&& \hspace{-4mm} -\frac{1}{4}F_{ij}^{ a}F_{ij}^{a} = -\frac{1}{4}\bigg[c^{2\lambda}\sum_{n,m=0}^{\infty}(\p_{i}a_{j}^{a(n)}-\p_{j}a_{i}^{a(n)})(\p_{i}a_{j}^{a(m)}-\p_{j}a_{i}^{a(m)}) +c^{3\lambda}\sum_{n,m,l=0}^{\infty}2gf^{abc}(\p_{i}a_{j}^{a{(n)}}\non\\
		&& \hspace{25mm}-\p_{j}a_{i}^{a(n)})a_{i}^{b(m)}a_{j}^{c(l)} 	+c^{4\lambda}\sum_{n,m,l,p=0}^{\infty}g^{2}f^{abc}f^{ade}a_{i}^{b(n)}a_{j}^{c(m)}a_{i}^{d(l)}a_{j}^{e(p)}\bigg].
	\end{eqnarray}
	
	We generally expand Lagrangian in even powers of $c$. If the relativistic action did not contain any self-interaction term, there would not have been any problem. We could have just taken $c^{\lambda}$ outside and written the Lagrangian in even power of $c$. But in our case, to write the expansions in Eq.\eqref{time part carroll expansion} and Eq.\eqref{Carroll expansion space part} in even powers of $c$, we have to choose $\lambda$ as an even integer. 
	We thus define  $\lambda=2\delta$. Then the two parts of the action become
	\bes\label{expansionofym} 
	\begin{eqnarray}
		&& \frac{1}{2c^{2}}F_{ti}^{ a}F_{ti}^{a}=\frac{1}{2}\bigg[c^{4\delta-2}\sum_{n,m=0}^{\infty}()+c^{6\delta-2}\sum_{n,m,l=0}^{\infty}()+c^{8\delta-2}\sum_{n,m,l,p=0}^{\infty}()\bigg], \\
		&& -\frac{1}{4}F_{ij}^{ a}F_{ij}^{a}=-\frac{1}{4}\bigg[c^{4\delta}\sum_{n,m=0}^{\infty}()+c^{6\delta}\sum_{n,m,l=0}^{\infty}()+c^{8\delta}\sum_{n,m,l,p=0}^{\infty}()\bigg],
	\end{eqnarray}
	\ees
	where $()$ is a shorthand for the corresponding terms in Eq.\eqref{time part carroll expansion} and Eq.\eqref{Carroll expansion space part}. Now every term looks good. As argued earlier, $\lambda = 0$ and $\lambda\neq 0$ (correspondingly $\delta=0$ and $\delta\neq 0$) give two distinct sectors. For $\delta=0$, the resultant Carrollian actions describe non-abelian theories, \ie\ these include the self-interaction terms, whereas for $\delta \neq 0$, the resultant Carrollian actions describe copies of the Carrollian abelian theory.
	
	\medskip
	
	\noindent For $\delta=0$, the leading order Lagrangian, \ie\ the coefficient of $c^{-2}$ in \eqref{expansionofym} is 
	\begin{eqnarray}\label{Electric00}&&
		\mathcal{L}^{(0)}=\frac{1}{2}\big[(\p_{t}a_{i}^{a(0)}-\p_{i}a_{t}^{a(0)})(\p_{t}a_{i}^{a(0)}-\p_{i}a_{t}^{a(0)})+2gf^{abc}(\p_{t}a_{i}^{a{(0)}}-\p_{i}a_{t}^{a(0)})a_{t}^{b(0)}a_{i}^{c(0)}\non\\&&\hspace{8cm}+g^{2}f^{abc}f^{ade}a_{t}^{b(0)}a_{i}^{c(0)}a_{t}^{d(0)}a_{i}^{e(0)}\big],
	\end{eqnarray}
	and this is called the electric sector. The next-to-leading order (NLO) Lagrangian (\ie\ the coefficient of $c^{0}$ in \eqref{expansionofym}), which is called the magnetic sector, is given by
		\begin{eqnarray}\label{magnetic00}
			&& \mathcal{L}^{(1)} = \big(\p_{t}a_{i}^{a(1)}-\p_{i}a_{t}^{a(1)}\big)E_{i}^{a(0)}+gf^{abc}\big(\p_{t}a_{i}^{a(0)}-\p_{i}a_{t}^{a(0)}\big)\big(a_{t}^{b(0)}a_{i}^{c(1)}+a_{t}^{b(1)}a_{i}^{c(0)}\big) \nonumber \\
			&& \hspace{12mm} +\frac{g^2}{2}f^{abc}f^{ade} \big[ a_{t}^{b(1)}a_{i}^{c(0)}a_{t}^{d(0)}a_{i}^{e(0)} + a_{t}^{b(0)}a_{i}^{c(1)}a_{t}^{d(0)}a_{i}^{e(0)} + a_{t}^{b(0)}a_{i}^{c(0)}a_{t}^{d(1)}a_{i}^{e(0)} \nonumber \\
			&& \hspace{34mm} + a_{t}^{b(0)}a_{i}^{c(0)}a_{t}^{d(0)}a_{i}^{e(1)} \big]-\frac{1}{4}f^{ija(0)}f_{ij}^{a(0)},
	\end{eqnarray}
	where $E_{i}^{a(0)}=\p_{t}a_{i}^{a(0)}-\p_{i}a_{t}^{a(0)}+gf^{abc}a_{t}^{a(0)}a_{i}^{a(0)}$ and  $f_{ij}^{a(0)}=\p_{i}a_{j}^{a(0)}-\p_{j}a_{i}^{a(0)}+gf^{abc}a_{i}^{a(0)}a_{j}^{a(0)}$.
	
	\medskip
	
	\noindent For $\delta\neq 0$, all values of $\delta$ are equivalent, and thus we take $\delta=1$ for simplicity. For $\delta = 1$, we get that the total Lagrangian in \eqref{expansionofym} has an expansion:
	$$\mathcal{L}=c^{2}\tilde{\mathcal{L}}_{0}+c^{4}\tilde{\mathcal{L}}_{1}+...,$$
	where the leading order Lagrangian (coefficient of $c^{2}$) is
	\begin{eqnarray}\label{11electric}
		\mathcal{\tilde{L}}^{(0)}=\frac{1}{2}(\p_{t}a_{i}^{a(0)}-\p_{i}a_{t}^{a(0)})(\p_{t}a_{i}^{a(0)}-\p_{i}a_{t}^{a(0)}),
	\end{eqnarray}
	and the next-to-leading order Lagrangian (coefficient of $c^{4}$) is
	\begin{eqnarray}\label{11nloaction}
		\mathcal{L}^{(1)}=\tilde{E}_{i}^{a(1)}\tilde{E}_{i}^{a(0)}+gf^{abc}\tilde{E}_{i}^{a(0)}a_{t}^{b(0)}a_{i}^{c(0)}-\frac{1}{4}\tilde{f}_{ij}^{a(0)}\tilde{f}_{ij}^{a(0)}.
	\end{eqnarray}
	Here $\tilde{E}^{a(0)}_{i}=(\p_{t}a_{i}^{a(0)}-\p_{i}a_{t}^{a(0)})$,  $\tilde{E}^{a(1)}_{i}=(\p_{t}a_{i}^{a(1)}-\p_{i}a_{t}^{a(1)})$ and  $\tilde{f}^{a(0)}_{ij}=(\p_{i}a_{j}^{a(0)}-\p_{j}a_{i}^{a(0)})$.
	
	\medskip
	
	\noindent Thus, taking $\lambda$ (\ie\ $\delta$) to be zero or non-zero, we have obtained four Lagrangians: two of these are the so-called electric sector, and the other two are the so-called magnetic sector. In the following sections, we will give details of all the four sectors.	
	
	\section{Carrollian Yang-Mills actions}\label{CYM all actions}
	\subsection{Electric Action I}\label{trivial electric}
	
	If we take $\delta=1$ in equation \eqref{expansionofym}, we can see the Leading order Lagrangian (coefficient of $c^{2}$) is given by
	\begin{eqnarray}\label{1electric}
		\mathcal{\tilde{L}}^{(0)}=\frac{1}{2}(\p_{t}a_{i}^{a(0)}-\p_{i}a_{t}^{a(0)})(\p_{t}a_{i}^{a(0)}-\p_{i}a_{t}^{a(0)})=\frac{1}{2}\tilde{E}_{i}^{a(0)}\tilde{E}_{i}^{a(0)},
	\end{eqnarray}
	where $\tilde{E}^{a(0)}_{i}=(\p_{t}a_{i}^{a(0)}-\p_{i}a_{t}^{a(0)})$. Unlike the $\delta=0$ case that we will study in the next subsection where the electric sector contains self-interaction, this just contains kinetic terms. The corresponding equations of motion are
	\begin{eqnarray}&&\label{EOM delta=1}
		\p_{i}\p_{t}a_{i}^{a(0)}-\p_{i}\p_{i}a_{t}^{a(0)}=\p_{i}\tilde{E}_{i}^{a}=0\label{1eom1},\quad\quad
		\p_{t}\p_{t}a_{i}^{a(0)}-\p_{t}\p_{i}a_{t}^{a(0)}=\p_{t}\tilde{E}_{i}^{a}=0.\label{1eom2}
	\end{eqnarray}
	The action and the equations of motion are copies of the electric sector of Carrollian abelian theory discussed in \cite{deBoer:2021jej}, where Carrollian symmetry (boost and rotation) is analyzed. Below, we will see the action's full infinite Carrollian conformal invariance. Boost and rotation invariance in our language are presented in Appendix \ref{app:rotation-boost}. 
	\subsection*{Gauge symmetry}
	The action here is just copies of the abelian action, so the gauge symmetry is like the abelian theory. The transformations are given by 
	\begin{eqnarray}&&
		a_{t}^{a(0)}\rightarrow a_{t}^{'a(0)}=a_{t}^{a(0)}+\p_{t}\alpha^{a},\quad
		a_{i}^{a(0)}\rightarrow a_{i}^{'a(0)}=a_{i}^{a(0)}+\p_{i}\alpha^{a}.
	\end{eqnarray}
	The action is invariant under the above gauge transformation, which are copies of the abelian gauge transformation. 
	\subsection*{Spacetime symmetries}
	 In the previous section, we talked about the gauge symmetry of the action. We will use the action of CCA to find the symmetries of the action Eq.\eqref{1electric}. In the representation theory sec.\ref{repcarroll}, we have some undefined constants. The value of these constants depends on the fields of the theory under consideration. For example, the value of scaling dimension ($\Delta$) for fields will be fixed when we impose dilatation invariance of the action. Similarly, all other constants will be fixed when we impose other symmetries of the action. All the four sectors of Lagrangian of the Carrollian Yang-Mills contain four sets of constants. Now let's discuss the first action we have stated above.
	
	\medskip
	
	The action is trivially invariant under time and space translations $(H,P_{i})$. The invariance of the action under rotation $(J_{ij})$, boost $(B_{i})$ are shown in Appendix \ref{app:rotation-boost}. Here we will only show the invariance under dilatation $(D)$, spatial special conformal transformation $(K_{i})$, and supertranslation $(M_{f})$. We know that for different values of $f$ the supertranslation $(M_{f})$ operator contains $B_{i}$ and $K$.

	\paragraph*{Dilatation:}
	Using the action of dilatation operator described in \eqref{repgen}, we write the transformations of fields under the dilatation operator. The transformations of $a_{t}^{a(0)}$ and $a_{i}^{a(0)}$ under dilatation is  
	\begin{eqnarray}\label{ed}
		\delta_{D}a_{t}^{a(0)}=(t\p_{t}+x^{k}\p_{k}+\Delta_{1})a_{t}^{a(0)},\quad\delta_{D}a_{i}^{a(0)}=(t\p_{t}+x^{k}\p_{k}+\Delta_{2})a_{i}^{a(0)}.
	\end{eqnarray}
	Using these transformations in action \eqref{1electric}, we can see the action changes as
	\begin{eqnarray}
		\delta_{D}\mathcal{L}=\p_{t}\big(t\tilde{E}_{i}^{a(0)}\tilde{E}_{i}^{a(0)}\big)+\p_{k}\big(x^{k}\tilde{E}_{i}^{a(0)}\tilde{E}_{i}^{a(0)}\big)\quad \text{if} \quad \Delta_{1}=\Delta_{2}=1.
	\end{eqnarray}
	So the action is invariant under dilatation in four spacetime dimensions if the scaling dimension is one for both the temporal and spatial components of gauge fields.
	
	\paragraph*{Spatial SCT:}
	Similar to the above case, the transformation of fields under spatial conformal transformation is given below. Here we take the transformation with arbitrary constants introduced when we discussed representation theory.  The transformations are given by
	\bes
	\begin{eqnarray}\label{eki}&&
		\delta_{K_{l}}a_{t}^{a(0)}=\big(2x_{l}+2x_{l}t\p_{t}+2x^{k}x_{l}\p_{k}-x_{k}x_{k}\p_{l}\big)a_{t}^{a(0)}+2tqa_{l}^{a(0)},\\&&
		\delta_{K_{l}}a_{i}^{a(0)}=\big(2x_{l}+2x_{l}t\p_{t}+2x^{k}x_{l}\p_{k}-x_{k}x_{k}\p_{l}\big)a_{i}^{a(0)}+2\delta_{li}x_{k}a_{k}^{a(0)}-2\delta_{lk}x_{i}a_{k}^{a(0)}+2tq'\delta_{li}a_{t}^{a(0)}.\non\\&&
	\end{eqnarray}
	\ees
	Using these transformations in action \eqref{1electric}, we can see the action changes as
	
	\begin{eqnarray}
		\delta_{K_{l}}\mathcal{L}^{(0)}=\p_{t}(x_{l}t\tilde{E}_{i}^{a(0)}\tilde{E}_{i}^{a(0)})+\p_{k}\big(x_{k}x_{l}\tilde{E}_{i}^{a}\tilde{E}_{i}^{a}\big)-\p_{l}\big(\frac{1}{2}x_{k}x_{k}\tilde{E}_{i}^{a(0)}\tilde{E}_{i}^{a(0)}\big).
	\end{eqnarray}
	So the action is invariant under spatial special conformal transformation if $q=0,\,q'=1$.
	
	\paragraph*{Supertranslation:}
	Instead of seeing the boost and temporal conformal transformation invariance of the action separately, we will see the supertranslation ($M_{f}$)  invariance of the action. 
	Fields transform under this operator($M_{f}$) as 
	\bes
	\begin{eqnarray}\label{emf}&&
		\delta_{M_{f}}a_{t}^{a(0)}=f(x)\p_{t}a_{t}^{a(0)},\quad
		\delta_{M_{f}}a_{i}^{a(0)}=f(x)\p_{t}a_{i}^{a(0)}+a_{t}^{a(0)}\p_{i}f(x).
	\end{eqnarray}
	\ees
	Using these transformations in action \eqref{1electric}, the action changes as
	
	\begin{eqnarray}
		\delta_{M_{f}}\mathcal{L}^{(0)}=\p_{t}\big(f(x)\tilde{E}_{i}^{a(0)}\tilde{E}_{i}^{a(0)}\big).
	\end{eqnarray}
	So the action is invariant under $M_{f}$. Thus we see that the action \eqref{1electric} is invariant under full infinite CCA in four spacetime dimensions.
	\subsection*{Energy-Momentum tensor}
	
	The components of the energy-momentum tensors for the action \eqref{1electric} are given by
	\begin{eqnarray}&&
		T^{t}\,_{t}=\tilde{E}_{i}^{a(0)}\p_{t}a_{i}^{a(0)}-\tilde{\mathcal{L}}^{(0)},\, T^{t}\,_{i}=\tilde{E}^{a(0)}_{j}\p_{i}a_{j}^{a(0)},\\&& T^{i}\,_{t}=-\tilde{E}^{a(0)}_{i}\p_{t}a_{t}^{a(0)},\,T^{i}_{j}=-E_{i}^{a(0)}\p_{j}a_{t}^{a(0)}-\delta^{i}_{j}\tilde{\mathcal{L}}^{(0)}.
	\end{eqnarray}
	Using the improvement of energy-momentum tensor defined in \cite{deBoer:2021jej}, the improved energy-momentum tensor in our case is
	\begin{eqnarray}\label{improve em for electric}
		T^{\mu}\,_{\nu}=-\frac{\delta\mathcal{L}}{\delta\p_{{\mu}}a_{\alpha}^{(0)a}}\p_{{\nu}}a_{\alpha}^{(0)a}+\delta^{\mu}\,_{\nu}\mathcal{L}-\big(\delta^{\mu}_{k}\p_{t}-\delta^{\mu}_{t}\p_{k}\big)\big(E^{(0)a}_{k}a_{\nu}^{(0)a}\big),
	\end{eqnarray}
	whose components are
	\begin{eqnarray}&&
		T^{t}\,_{i}=\tilde{E}_{j}^{a(0)}f_{ij}^{a(0)},\quad T^{i}\,_{t}=0, \quad T^{t}\,_{t}=\frac{1}{2}\tilde{E}_{j}^{a(0)}\tilde{E}_{j}^{a(0)},\\&& T^{i}\,_{j}=\tilde{E}_{i}^{a(0)}\tilde{E}_{j}^{a(0)}-\frac{1}{2}\delta^{i}_{j}\tilde{E}_{k}^{a(0)}\tilde{E}^{a(0)}_{k}.
	\end{eqnarray}
	We can see the energy-momentum tensor is gauge invariant, traceless, and symmetric under the interchange of spatial indices. The $T^{i}\,_{t}$ component of the stress tensor is zero as required by the Carroll symmetries. Below, we will see the conservation of energy-momentum tensor.
	
	\medskip
	
	The relativistic Bianchi identity for Yang-Mills is given in Eq.\eqref{rbianchi}. Now we are considering Carrollian Yang-Mills theory with $\delta=1$, and the Carrollian Bianchi identities for this case are
	\bes\label{1bianchi}
	\begin{eqnarray}&&
		\p_{t}\tilde{f}_{jk}^{a}+\p_{j}\tilde{f}_{kt}^{a}+\p_{k}\tilde{f}_{tj}^{a}=0\label{1bi1},\\&&
		\p_{i}\tilde{f}_{jk}^{a}+\p_{j}\tilde{f}_{ki}^{a}+\p_{k}\tilde{f}_{ij}^{a}=0\label{1bi2}.
	\end{eqnarray}
	\ees
	Tilde means there are no interaction terms in the field strength, only abelian terms.
	These are just copies of the Carrollian abelian Bianchi identity discussed in\cite{deBoer:2021jej}.
	Using equations of motion and Carrollian Bianchi identity we can see
	\bes
	\begin{eqnarray}\label{cem}&&
		\p_{t}T^{t}\,_{t}+\p_{i}T^{i}_{t}=0,\quad using \quad \eqref{1eom1},\eqref{1bi1},\\&&
		\p_{t}T^{t}_{j}+\p_{i}T^{i}_{j}=0,\quad using \quad \eqref{1eom2},\eqref{1bi2}.
	\end{eqnarray}
	\ees
	So the energy-momentum tensors satisfies conservation equations. We will return to this section when discussing Noether charges and Quantum aspects.
	
\subsection{Electric Action II}\label{Non-trivial electric}

The electric sector action Eq.\eqref{Electric00}, which has a non-abelian term, can be written in compact form:
\begin{eqnarray}\label{Electric0}&&
	\mathcal{L}_{0}=\frac{1}{2}\bigg((\p_{t}a_{i}^{a(0)}-\p_{i}a_{t}^{a(0)})(\p_{t}a_{i}^{a(0)}-\p_{i}a_{t}^{a(0)})+2gf^{abc}(\p_{t}a_{i}^{a{(0)}}-\p_{i}a_{t}^{a(0)})a_{t}^{b(0)}a_{i}^{c(0)}\non\\&&\hspace{5.1cm}+g^{2}f^{abc}f^{ade}a_{t}^{b(0)}a_{i}^{c(0)}a_{t}^{d(0)}a_{i}^{e(0)}\bigg)=\frac{1}{2}E_{i}^{a(0)}E_{i}^{a(0)},
\end{eqnarray}
where $E_{i}^{a(0)}=\p_{t}a_{i}^{a(0)}-\p_{i}a_{t}^{a(0)}+gf^{abc}a_{t}^{a(0)}a_{i}^{a(0)}$. The equations of motion following from the action are given by
\bes\label{EOM}
\begin{eqnarray}&&
	\p_{i}E_{i}^{a(0)}+gf^{abc}a_{i}^{b(0)}E_{i}^{c(0)}=D_{i}^{(0)}E_{i}^{a(0)}=0\label{0eeom1},\\&&
	\p_{t}E_{i}^{a(0)}+gf^{abc}a_{t}^{b(0)}E_{i}^{c(0)}=D_{t}^{(0)}E_{i}^{a(0)}=0\label{0eeom2},
\end{eqnarray}
\ees
where $D_{i}\mathcal{O}^{a}=\p_{i}\mathcal{O}^{a}+gf^{abc}a_{i}^{b(0)}\mathcal{O}^{c},\, D_{t}\mathcal{O}^{a}=
\p_{t}\mathcal{O}^{a}+gf^{abc}a_{t}^{b(0)}\mathcal{O}^{c}$.
\medskip

\subsection*{Gauge Symmetry} 
The gauge transformations under which the action \eqref{Electric0} is invariant are given by
\bes\label{electric gauge symmetry}
\begin{eqnarray}&&
	a_{t}^{a(0)}\rightarrow a_{t}^{a(0)'}=a_{t}^{a(0)}+\frac{1}{g}\p_{t}\alpha^{a}+f^{abc}a_{t}^{b(0)}\alpha^{c},\\&&
	a_{i}^{a(0)}\rightarrow a_{i}^{a(0)'}=a_{i}^{a(0)}+\frac{1}{g}\p_{i}\alpha^{a}+f^{abc}a_{i}^{b(0)}\alpha^{c}.
\end{eqnarray}
\ees
This gauge transformation is the same as parent theory, but now we cannot write it in covariant form like relativistic theory. Because, like the non-relativistic theory, the metrics in Carrollian theory are degenerate, and time and space are not on the same footing.

\subsection*{Spacetime Symmetries}


\paragraph*{Dilatation:}
Using the action of dilatation operator described in Eq.~\eqref{repgen}, we write the transformations of fields under the dilatation operator. The transformations of $a_{t}^{a(0)}$ and $a_{i}^{a(0)}$ under dilatation is  
\begin{eqnarray}\label{ed}
	\delta_{D}a_{t}^{a(0)}=(t\p_{t}+x^{k}\p_{k}+\Delta_{1})a_{t}^{a(0)},\quad\delta_{D}a_{i}^{a(0)}=(t\p_{t}+x^{k}\p_{k}+\Delta_{2})a_{i}^{a(0)}.
\end{eqnarray}
Using these transformations in action \eqref{Electric0}, we can see the action changes as
\begin{eqnarray}
	\delta_{D}\mathcal{L}=\p_{t}\big(tE_{i}^{a(0)}E_{i}^{a(0)}\big)+\p_{k}\big(x^{k}E_{i}^{a(0)}E_{i}^{a(0)}\big)\quad \text{if} \quad \Delta_{1}=\Delta_{2}=1\, .
\end{eqnarray}

If the scaling dimensions of both fields $\Delta_{1}$ and $\Delta_{2}$ are one, then the action is dilatation invariant in four spacetime dimensions.

\paragraph*{Spatial SCT:}
Similar to the above case, the transformation of fields under spatial conformal transformation is given below. Here we take transformation with arbitrary constant introduced when we discussed representation theory. The transformations are given by
\bes
\begin{eqnarray}\label{eki}&&
	\delta_{K_{l}}a_{t}^{a(0)}=\big(2x_{l}+2x_{l}t\p_{t}+2x^{k}x_{l}\p_{k}-x_{k}x_{k}\p_{l}\big)a_{t}^{a(0)}+2tqa_{l}^{a(0)},\\&&
	\delta_{K_{l}}a_{i}^{a(0)}=\big(2x_{l}+2x_{l}t\p_{t}+2x^{k}x_{l}\p_{k}-x_{k}x_{k}\p_{l}\big)a_{i}^{a(0)}+2\delta_{li}x_{k}a_{k}^{a(0)}-2\delta_{lk}x_{i}a_{k}^{a(0)}+2tq'\delta_{li}a_{t}^{a(0)}.\non\\&&
\end{eqnarray}
\ees
Using these transformations in action \eqref{Electric0}, we can see the action changes as
\begin{eqnarray}
	\delta_{K_{l}}\mathcal{L}^{(0)}=\p_{t}(x_{l}tE_{i}^{a(0)}E_{i}^{a(0)})+\p_{k}\big(x_{k}x_{l}E_{i}^{a}E_{i}^{a}\big)-\p_{l}\big(\frac{1}{2}x_{k}x_{k}E_{i}^{a(0)}E_{i}^{a(0)}\big)\\\text{if} \quad q=0,\,q'=1,
\end{eqnarray}
so here, we can see that the action is spatial special conformal invariant if the constants $q$ and $q^{'}$, respectively, are zero and one.
\paragraph*{Supertranslation:}
Instead of seeing the boost and temporal conformal transformation invariance of the action separately, we will see the supertranslation ($M_{f}$)  invariance of the action. 
Fields transform under this operator($M_{f}$) as 
\bes
\begin{eqnarray}\label{emf}&&
	\delta_{M_{f}}a_{t}^{a(0)}=f(x)\p_{t}a_{t}^{a(0)},\quad
	\delta_{M_{f}}a_{i}^{a(0)}=f(x)\p_{t}a_{i}^{a(0)}+a_{t}^{a(0)}\p_{i}f(x).
\end{eqnarray}
\ees
Using these transformations in action \eqref{Electric0}, the action changes as
\begin{eqnarray}
	\delta_{M_{f}}\mathcal{L}^{(0)}=\p_{t}\big(f(x)E_{i}^{a(0)}E_{i}^{a(0)}\big),
\end{eqnarray}
so the action is invariant under supertranslation ($M_{f}$).

So from the above analysis, we conclude that the action \eqref{Electric0} is invariant under infinite CCA in spacetime dimension four if the scaling dimensions for both $a_{t}^{a}$ and $a_{i}^{a}$ are one.

\subsection*{Energy-Momentum tensor}
The leading order Lagrangian or so-called electric sector is infinite Carrollian conformal invariant in $4d$ spacetime. Now let's see the energy-momentum tensor for the action Eq.\eqref{Electric0} and how we can improve it.
Different components of energy-momentum tensor for action \eqref{Electric0} are given by
\begin{eqnarray}&&
	T^{t}\,_{i}=E_{j}^{a(0)}\p_{i}a_{j}^{a(0)},\quad T^{i}\,_{t}=-E^{a(0)}_{i}\p_{t}a_{t}^{a(0)}, \quad T^{t}\,_{t}=E_{i}^{a(0)}\p_{t}a_{i}^{a(0)}-\mathcal{L}^{(0)},\\&& \hspace{3cm}T^{i}\,_{j}=-E_{i}^{a(0)}\p_{j}a_{t}^{a}-\delta^{i}_{j}\mathcal{L}^{(0)}.
\end{eqnarray}
We can see these are not gauge invariant, $T^{i}\,_{j}$ component is not symmetric, and $T^{i}\,_{t}$ component is not zero, so we have to improve it.  
Using the improved energy-momentum tensor defined in \cite{deBoer:2021jej}, the improved energy-momentum tensor for our case is
\begin{eqnarray}\label{improve em for electric}
	T^{\mu}\,_{\nu}=-\frac{\delta\mathcal{L}}{\delta\p_{{\mu}}a_{\alpha}^{(0)a}}\p_{{\nu}}a_{\alpha}^{(0)a}+\delta^{\mu}\,_{\nu}\mathcal{L}-\big(\delta^{\mu}_{k}\p_{t}-\delta^{\mu}_{t}\p_{k}\big)\big(E^{(0)a}_{k}a_{\nu}^{(0)a}\big),
\end{eqnarray}
and the components of improved E-M tensor are
\begin{eqnarray}&&
	T^{t}\,_{i}=E_{j}^{a(0)}f_{ij}^{a(0)},\quad T^{i}\,_{t}=0, \quad T^{t}\,_{t}=\frac{1}{2}E_{j}^{a(0)}E_{j}^{a(0)},\\&& T^{i}\,_{j}=E_{i}^{a(0)}E_{j}^{a(0)}-\frac{1}{2}\delta^{i}_{j}E_{k}^{a(0)}E^{a(0)}_{k}.
\end{eqnarray}
We can see the energy-momentum tensor is gauge invariant, traceless, and symmetric under the interchange of spatial indices. The $T^{i}\,_{t}$ component of the stress tensor is zero as required by Carroll symmetries. Below, we will see the conservation of energy-momentum tensor.

\medskip

Before going to the conservation of energy-momentum tensor, let's see Bianchi's identity in the Carrollian limit. The relativistic Bianchi identity for the Yang-Mills is given by
\begin{eqnarray}\label{rbianchi}
	D_{\nu}F_{\beta\mu}^{a}+D_{\beta}F_{\mu\nu}^{a}+D_{\mu}F_{\nu\beta}^{a}=0.
\end{eqnarray}
When discussing the expansion of action, we discussed the different values of $\lambda$. For different $\lambda$, we will also have two distinct Bianchi identities. For $\lambda=0$ or $(\delta=0)$, we have Bianchi identity with a non-trivial self-interaction term. And for $\lambda=2$ or $(\delta=1)$, we will have Bianchi identity copies of Carrollian abelian Bianchi identity.  Which we have mentioned in the previous sector.
For $\lambda=0(\delta=0)$, {Carrollian Bianchi identity} is 
\bes\label{cbi}
\begin{eqnarray}&&
	D_{t}f_{jk}^{a}+D_{j}f_{kt}^{a}+D_{k}f_{tj}^{a}=0\label{0bi1},\\&&
	D_{i}f_{jk}^{a}+D_{j}f_{ki}^{a}+D_{k}f_{ij}^{a}=0\label{0bi2}.
\end{eqnarray}
\ees
Using equations of motion and above Carrollian Bianchi identity \eqref{cbi}, we can see the conservation of energy-momentum tensor as
\begin{eqnarray}\label{cem}&&
	\p_{t}T^{t}\,_{t}+\p_{i}T^{i}\,_{t}=0,\quad \text{using}\quad\eqref{0eeom2},\\&&
	\p_{t}T^{t}\,_{j}+\p_{i}T^{i}\,_{j}=0, \quad \text{using}\quad \eqref{0eeom1},\eqref{0bi1}.
\end{eqnarray}
This is our detailed discussion on the electric sector with non-abelian terms. In the next section, we will focus on the magnetic sector. In the electric sector, we have seen that the temporal components of field strength are dominated; in the subsequent section in the magnetic sector, we will see the purely spatial sector of field strength will dominate, and the temporal component will behave as a constraint. We will visit this electric sector again when we discuss Noether charges and quantum aspects of the theory.
	
	\subsection{Magnetic Action I}\label{trivial magnetic}
	For $\delta=1$ case, the next to leading order(NLO) Lagrangian or the so called magnetic sector is given in Eq.\eqref{11nloaction}.  For convenience, let's write the action again here
	\begin{eqnarray}\label{1nloaction}
		\mathcal{L}^{(1)}=\tilde{E}_{i}^{a(1)}\tilde{E}_{i}^{a(0)}+gf^{abc}\tilde{E}_{i}^{a(0)}a_{t}^{b(0)}a_{i}^{c(0)}-\frac{1}{4}\tilde{f}_{ij}^{a(0)}\tilde{f}_{ij}^{a(0)}.
	\end{eqnarray}
	
	If we take the variation of the Lagrangian with respect to next to leading order fields $a_{t}^{a(1)},a_{i}^{a(1)}$ we will get  Eq.\eqref{EOM delta=1}, the leading order equation of motion. If we take variation of the action w.r.t  $a_{t}^{a(0)},a_{i}^{a(0)}$ (the leading order fields) the equations of motion are
	\bes\label{1nloeom}
	\begin{eqnarray}&&
		\p_{i}\tilde{E}_{i}^{a(1)}+gf^{abc}\p_{i}\big(a_{t}^{b(0)}a_{i}^{c(0)}\big)+gf^{abc}a_{i}^{b(0)}\tilde{E}_{i}^{c(0)}=0,\\&&
		\p_{t}\tilde{E}_{i}^{a(1)}+gf^{abc}\p_{t}\big(a_{t}^{b(0)}a_{i}^{c(0)}\big)+gf^{abc}a_{t}^{b(0)}\tilde{E}_{i}^{c(0)}-\p_{k}\tilde{f}_{ki}^{a(0)}=0.
	\end{eqnarray}
	\ees
	The action Eq.\eqref{1nloaction} and the above equations of motion are not Carroll invariant. To make these Carroll invariant, we have to impose constraints $\tilde{E}_{i}^{a(0)}=0$ in action \eqref{1nloaction}. The corresponding equation of motion is 
	\begin{eqnarray}
		\p_{k}\tilde{f}_{ki}^{a(0)}=0.
	\end{eqnarray}

	Similar to $\delta=0$ case, we will derive the Carrollian invariant magnetic sector for the $\delta=1$ case using the Lagrange multiplier in the parent action. We can start from relativistic Lagrangian with Lagrange multiplier $\xi_{i}^{a}$,
	\begin{eqnarray}
		\mathcal{L}=-\frac{c^{2}}{2}\xi_{i}^{a}\xi_{i}^{a}+\xi_{i}^{a}F_{0i}^{a}-\frac{1}{4}F_{ij}^{a}F_{ij}^{a}.
	\end{eqnarray}
	From equation Eq.\eqref{carroll expansion of fields}, we can see that for $\delta=1$ case, before doing the expansion, every field gets scaled by $c^{2}$. If we scale every field by $c^{2}$ of the above equation and collect the $c^{4}$ term (for $\delta=1$ case, NLO action is of the order of $c^{4}$ term of expansion \eqref{expansionofym}), the resultant Lagrangian is
	\begin{eqnarray}\label{1magaction}
		\mathcal{\tilde{L}}^{NLO}=\xi_{i}^{a}\tilde{E}_{i}^{a}-\frac{1}{4}\tilde{f}_{ij}^{a}\tilde{f}_{ij}^{a}.
	\end{eqnarray}
	If we vary the Lagrangian with respect to $\xi_{i}^{a}$, we will get $\tilde{E}_{i}^{a}=0$ constraint. So all the equations of motion of the Lagrangian are 
	\begin{eqnarray}\label{1meom}&&
		\tilde{E}_{i}^{a}=0,\quad \p_{i}\xi_{i}^{a}=0,\quad \p_{t}\xi_{i}^{a}-\p_{j}\tilde{f}_{ji}^{a}=0.
	\end{eqnarray}
	
	We will see its gauge symmetry and full spacetime symmetry below. 
	\subsection*{Gauge symmetry} 
	 The gauge symmetry of the action is not non-abelian. It reduces to copies of abelian or $u(1)$ symmetry.
	The action is invariant under the gauge transformation
	\begin{eqnarray}&&
		a_{t}^{a}\rightarrow a_{t}^{'a}=a_{t}^{a}+\p_{t}\alpha^{a},\quad
		a_{i}^{a}\rightarrow a_{i}^{'a}=a_{i}^{a}+\p_{i}\alpha^{a},\quad
		\xi_{i}^{a}\rightarrow \xi_{i}^{'a}=\xi_{i}^{a}.
	\end{eqnarray}
	So the action is symmetric under $n^{2}-1$ copies of abelian symmetry. The Lagrange multiplier $\xi_{i}^{a}$ behaves as a scalar under gauge transformation.
	\subsection*{Spacetime symmetries}
	The action \eqref{1magaction} is copies of the magnetic sector of Carrollian abelian theory discussed in \cite{deBoer:2021jej}. Carroll symmetry of the action is analyzed in that paper. Here we will see Carrollian conformal invariance of the action. The transformations of different fields are the same as magnetic sector fields of $\delta=0$. In this section, we only give how action changes under dilatation, spatial SCT, and supertranslation. If readers want to see rotation and boost invariance of action, they can check the appendix.\paragraph*{Dilatation:}
	Transformations of gauge fields($a_{t}^{a},a_{i}^{a}$) and Lagrange multiplier($\xi_{i}^{a}$) under the dilatation operator($D$) are  given by
	\bes
	\begin{eqnarray}\label{0magD}&&
		\delta_{D}a_{t}^{a}=(t\p_{t}+x^{k}\p_{k}+\Delta_{1})a_{t}^{a},\quad\delta_{D}a_{i}^{a}=(t\p_{t}+x^{k}\p_{k}+\Delta_{2})a_{i}^{a},\\&&
		\hspace{3cm}{\delta_D}\xi_{i}^{a}=(t\p_{t}+x^{k}\p_{k}+\Delta_{\xi})\xi_{i}^{a}.
	\end{eqnarray}
	\ees
	Using this transformation in \eqref{1magaction}, the change of action as
	
	\begin{eqnarray}&&
		\delta_{D}\mathcal{L}=\p_{t}\big(t\tilde{E}_{i}^{a}\tilde{E}_{i}^{a}\big)+\p_{k}\big(x^{k}\tilde{E}_{i}^{a}\tilde{E}_{i}^{a}\big)+\p_{t}\big(-\frac{1}{4}\tilde{f}_{ij}^{a}\tilde{f}_{ij}^{a}\big)+\p_{k}\big(-\frac{1}{4}\tilde{f}_{ij}^{a}\tilde{f}_{ij}^{a}\big)\non\\&&
		\hspace{5cm}\text{if} \quad \Delta_{1}=\Delta_{2}=1 \quad \text{and}\quad \Delta_{\xi}=2,
	\end{eqnarray}
	so the action is invariant under dilatation transformation in four spacetime dimensions if the scaling dimensions of the temporal and spatial component of gauge fields are one, and for $\xi_{i}$ scaling dimension is two.
	
		\paragraph*{Spatial SCT:}
	Transformation of fields $a_{t}^{a}$, $a_{i}^{a}$ and $\xi_{i}^{a}$ under spatial SCT  are given by
	\bes\label{0magKi}
	\begin{eqnarray}&&
		\hspace{-1cm}\delta_{K_{l}}a_{t}^{a}=\big(2x_{l}+2x_{l}t\p_{t}+2x^{k}x_{l}\p_{k}-x_{k}x_{k}\p_{l}\big)a_{t}^{a}+2tqa_{l}^{a},\\&&
		\hspace{-1cm}\delta_{K_{l}}a_{i}^{a}=\big(2x_{l}+2x_{l}t\p_{t}+2x^{k}x_{l}\p_{k}-x_{k}x_{k}\p_{l}\big)a_{i}^{a}+2\delta_{li}x_{k}a_{k}^{a}-2\delta_{lk}x_{i}a_{k}^{a}+2tq'\delta_{li}a_{t}^{a},\\&&
		\hspace{-1cm}\delta_{K_{l}}\xi_{i}^{a}=\big(4x_{l}+2x_{l}t\p_{t}+2x^{k}x_{l}\p_{k}-x_{k}x_{k}\p_{l}\big)\xi_{i}^{a}+2\delta_{li}x_{k}\xi_{k}^{a}-2\delta_{lk}x_{i}\xi_{k}^{a}+2tq''\delta_{li}a_{t}^{a}\non\\&&\hspace{10.5cm}+2q'''t\tilde{f}_{il}^{a},
	\end{eqnarray}
	\ees
	using these transformations in \eqref{1magaction}, the action changes as
	\begin{eqnarray}&&
		\delta_{K_{l}}\mathcal{L}^{(0)}=\p_{t}(2x_{l}t\xi_{i}^{a}\tilde{E}_{i}^{a(0)})-\p_{t}\big(\frac{1}{2}tx_{l}\tilde{f}_{ij}^{a}\tilde{f}_{ij}^{a}\big)+\p_{k}\big(2x_{k}x_{l}\xi_{i}^{a}\tilde{E}_{i}^{a}\big)-\p_{k}\big(\frac{1}{2}x_{k}x_{l}\tilde{f}_{ij}^{a}\tilde{f}_{ij}^{a}\big)\non\\&&\hspace{.5cm}-\p_{l}\big(x_{k}x_{k}\xi_{i}^{a}\tilde{E}_{i}^{a(0)}\big)+\p_{l}\big(\frac{1}{4}x_{k}x_{k}\tilde{f}_{ij}^{a}\tilde{f}_{ij}^{a}\big), \, \text{if} \,\, q=0,\,q'=1,\,q''=0,\,q'''=-1.
	\end{eqnarray}
	So the action is invariant under spatial conformal transformation.	\paragraph*{Supertranslation: }
	Lastly, invariance under supertranslation ($M_{f}$), which contains Hamiltonian, temporal spacial conformal, and boost operator for different choice of $f$. Under this operator, fields transform as
	\begin{eqnarray}\label{0magMf}&&
		\hspace{-.9cm}\delta_{M_{f}}a_{t}^{a}=f(x)\p_{t}a_{t}^{a},\quad
		\delta_{M_{f}}a_{i}^{a}=f(x)\p_{t}a_{i}^{a}-a_{t}^{a}\p_{i}f(x),\quad
		\delta_{M_{f}}\xi_{i}^{a}=f(x)\p_{t}\xi_{i}^{a}-f_{ik}^{a}\p_{k}f(x),
	\end{eqnarray}
	using these in \eqref{1magaction}, the actions changes as
	\begin{eqnarray}
		\delta_{M_{f}}\mathcal{L}^{(0)}=\p_{t}\big(f(x)\xi_{i}^{a}\tilde{E}_{i}^{a(0)}-\frac{1}{4}\tilde{f}_{ij}^{a}\tilde{f}_{ij}^{a}\big)+\p_{i}\big(-\frac{1}{2}\tilde{f}_{ij}^{a}\tilde{E}_{j}^{a}\big)+\p_{j}\big(\frac{1}{2}\tilde{f}_{ij}^{a}\tilde{E}_{i}^{a}\big).
	\end{eqnarray}
	So the action is invariant under $M_{f}$. The magnetic sector with $\delta=1$  case is invariant under infinite CCA in $4d$ spacetime. 
	\subsection*{Energy-Momentum tensor}
	Now we will see the energy-momentum tensor and its conservation. If we derive the energy-momentum tensor from \eqref{1magaction} we will get
	\begin{eqnarray}&&
		T^{t}\,_{i}=\xi_{k}^{a}\p_{i}a_{k}^{a},\quad T^{i}\,_{t}=-x_{i}^{a}\p_{t}a_{t}^{a}-\tilde{f}_{ik}^{a}\p_{t}a_{k}^{a}\\&&
		T^{t}\,_{t}=\xi_{i}^{a}\p_{t}a_{k}^{a}-\mathcal{L},\quad T^{i}\,_{j}=-\xi_{i}^{a}\p_{j}a_{t}^{a}-\tilde{f}_{ik}^{a}\p_{j}a_{k}^{a}-\delta^{i}_{j}\mathcal{L}.
	\end{eqnarray}
	Here we also need an improved energy-momentum tensor as the electric sector. Following \cite{deBoer:2021jej},
	the improved energy-momentum tensor formula for the magnetic sector is
	\begin{equation}\label{improve em magnetic}
		T^{\mu}{_\nu}
		=
		-
		\frac{\delta\mathcal{L}}{\delta\partial_{\mu}a_{\alpha}^{a}}\partial_{\nu}a_{\alpha}^{a}
		+
		\delta^{\mu}{_\nu}\mathcal{L}
		-
		\delta^{\mu}{_t}
		\partial_{i}
		\left[
		\xi^{a}_{i}a_{\nu}^{a}
		\right]
		+
		\delta^{\mu}{_i}
		\left[
		\partial_{t}(\xi_{i}^{a}a_{\nu}^{a})
		+
		\partial_{j}(\tilde{f}_{ij}^{a}a_{\nu}^{a})
		\right]
		\,.
	\end{equation} 
	If we write it components wise explicitly
	\begin{eqnarray}&&
		T^{t}\,_{i}=\xi_{k}^{a}\tilde{f}_{ik}^{a},\quad T^{i}\,_{t}=0,\quad
		T^{t}\,_{t}=\frac{1}{4}\tilde{f}_{ij}^{a}\tilde{f}_{ij}^{a},\quad T^{i}\,_{j}=-\tilde{f}_{ik}^{a}\tilde{f}_{jk}^{a}-\delta^{i}_{j}\mathcal{L}.
	\end{eqnarray}
	Here we can see energy-momentum tensor is gauge invariant, traceless, symmetric in spatial indices, and $T^{i}_{t}=0$ as expected for Carroll theory. Using equations of motion and Carrollian Bianchi identity, we can see
	\bes
	\begin{eqnarray}\label{cem}&&
		\p_{t}T^{t}\,_{t}+\p_{i}T^{i}\,_{t}=0,\quad \text{using} \quad \eqref{1meom},\eqref{0bi1}.\\&&
		\p_{t}T^{t}\,_{j}+\p_{i}T^{i}\,_{j}=0,\quad\text{using} \quad \eqref{1meom},\eqref{0bi2}.
	\end{eqnarray}
	\ees
	the energy-momentum tensor satisfies conservation equations.
	
	
	\medskip
	\subsection{Magnetic Action II}\label{non-trivial magnetic}
	
	In this section, we will study details of the NLO Lagrangian or so-called magnetic sector. The NLO Lagrangian contains leading order and NLO fields. 
	From the expansion of action section, we have the NLO Lagrangian (coefficient of $c^{0}$) in Eq.\eqref{magnetic00}.
	The action looks horrible to analyze. Thanks to Jacobi's identity, 
	\begin{eqnarray}\label{jacobi}
		f_{bca}f_{dae}+f_{dba}f_{cae}+f_{cda}f_{bae}=0,
	\end{eqnarray}
	using this, we can simplify the action.
	Using the above identity and doing some calculations, we can write the NLO Lagrangian in this form
	\begin{eqnarray}\label{wron0mag}
		\mathcal{L}^{(1)}=\big(D_{t}^{(0)}a_{i}^{a(1)}\big)E_{i}^{a(0)}-\big(D_{i}^{(0)}a_{t}^{a(1)}\big)E_{i}^{a(0)}-\frac{1}{4}f^{ija(0)}f_{ij}^{a(0)}.
	\end{eqnarray}
	If we take the variation of the Lagrangian with respect to next to leading order fields $a_{t}^{a(1)},a_{i}^{a(1)}$ we will get  Eq.\eqref{EOM}, leading order equations of motion as a property of this formalism. If we take variation with respect to leading order fields ($a_{t}^{a(0)},a_{i}^{a(0)}$), equations of motion are
	\bes\label{NLO EOM}
	\begin{eqnarray}&&
		D_{i}^{(0)}D_{i}^{(0)}a_{t}^{a(1)}-D_{i}^{(0)}D_{t}^{(0)}a_{i}^{a(1)}-gf^{abc}a_{i}^{b(1)}E_{i}^{c(0)}=0,\\&&
		D_{t}^{(0)}D_{t}^{(0)}a_{i}^{a(1)}-D_{t}^{(0)}D_{i}^{(0)}a_{t}^{a(1)}-gf^{abc}a_{t}^{b(1)}E_{i}^{c(0)}-D_{k}^{(0)}f_{ki}^{a(0)}=0,
	\end{eqnarray}
	\ees
	where $D_{k}^{(0)}f_{ki}^{a(0)}=\p_{k}f_{ki}^{a(0)}+gf^{abc}a_{k}^{b(0)}f_{ki}^{c(0)}$.
	Although the action and the equations of motion look nice in compact form, these are not Carroll invariant. To make Carroll invariant, we have to take the constraint $E_{i}^{a(0)}=0$ at the level of action Eq.\eqref{wron0mag}. Then action will become  $-\frac{1}{4}f^{ija(0)}f_{ij}^{a(0)}$ and equations of motion will be $D_{k}^{(0)}f_{ki}^{a(0)}=0$. 
	
	We can derive the Carroll invariant magnetic sector from the Relativistic Yang-Mills action if we consider a Lagrange multiplier in relativistic Lagrangian and then take speed of light to zero limit. The relativistic Lagrangian with Lagrange multiplier $\xi_{i}^{a}$ and explicit $c$ factor is given by
	\begin{eqnarray}
		\mathcal{L}=-\frac{c^{2}}{2}\xi_{i}^{a}\xi_{i}^{a}+\xi_{i}^{a}F_{0i}^{a}-\frac{1}{4}F_{ij}^{a}F_{ij}^{a}.
	\end{eqnarray}
	From here, we can get back to the usual Yang-Mills action if we integrate out $\xi_{i}$ fields. Now we can see if we take the small $c$ limit here, we will get
	\begin{eqnarray}\label{omag}&&
		\mathcal{L}^{NLO}=\xi_{i}^{a}(\p_{t}a_{i}^{a(0)}-\p_{i}a_{t}^{a(0)})-
		\frac{1}{4}(\p_{i}a_{j}^{a}-\p_{j}a_{i}^{a})(\p_{i}a_{j}^{a}-\p_{j}a_{i}^{a})+
		gf^{abc}a_{t}^{b}a_{i}^{c}\xi_{i}^{a}\non\\&&\hspace{3cm}-gf^{abc}a_{i}^{b}a_{j}^{c}\p_{i}a_{j}^{a}-\frac{1}{4}g^{2}f^{abc}f^{ade}a_{i}^{b}a_{j}^{c}a_{i}^{d}a_{j}^{e}
		=\xi_{i}^{a}E_{i}^{a}-\frac{1}{4}f_{ij}^{a}f_{ij}^{a}.
	\end{eqnarray}
	The Lagrangian contains non-trivial self-interaction terms or non-abelian terms.  The equations of motion of this action are
	\begin{eqnarray}\label{0meom}&&
		E_{i}^{a}=0,\quad D_{i}\xi_{i}^{a}=0,\quad D_{t}\xi_{i}-D_{j}f_{ji}=0.
	\end{eqnarray}
	Here we are getting the constraints $E_{i}^{a(0)}=0$ as an equations of motion for the Lagrange($\xi_{i}^{a}$). Below we will see the full spacetime symmetry of this action.
	
	\medskip
	\subsection*{Gauge symmetry}
	Before seeing the spacetime symmetry of the action, it will be good to check the gauge symmetry. The action Eq.\eqref{omag} is invariant under the gauge transformation
	\bes
	\begin{eqnarray}&&
		a_{t}^{a}\rightarrow a_{t}^{'a}=a_{t}^{a}+\frac{1}{g}\p_{t}\alpha^{a}+f^{abc}a_{t}^{b}\alpha^{c},\\&&
		a_{i}^{a}\rightarrow a_{i}^{'a}=a_{i}^{a}+\frac{1}{g}\p_{i}\alpha^{a}+f^{abc}a_{i}^{b}\alpha^{c},\\&&
		\xi_{i}^{a}\rightarrow \xi_{i}^{'a}=\xi_{i}^{a}+f^{abc}\xi_{i}^{b}\alpha^{c}.
	\end{eqnarray}
	\ees
	The temporal and spatial component of the gauge field is transformed in the same way as the electric sector. The Lagrange multiplier $\xi_{i}^{a}$ transforms as a scalar in the adjoint representation of the underlying gauge group.
	\subsection*{Spacetime symmetries}
	Similar to the electric sector discussed above, we will see the symmetry of the action under dilatation ($D$), spatial SCT ($K_{i}$), and supertranslation $M_{f}$.
	The Rotation and boost invariance is shown in the appendix. 
	\paragraph*{Dilatation:}
	Transformations of gauge fields($a_{t}^{a},a_{i}^{a}$) and Lagrange multiplier($\xi_{i}^{a}$) under the dilatation operator($D$) are  given by
	\bes
	\begin{eqnarray}\label{0magD}&&
		\delta_{D}a_{t}^{a}=(t\p_{t}+x^{k}\p_{k}+\Delta_{1})a_{t}^{a},\quad\delta_{D}a_{i}^{a}=(t\p_{t}+x^{k}\p_{k}+\Delta_{2})a_{i}^{a},\\&&
		\hspace{3cm}{\delta_D}\xi_{i}^{a}=(t\p_{t}+x^{k}\p_{k}+\Delta_{\xi})\xi_{i}^{a},
	\end{eqnarray}
	\ees
	using this transformation in \eqref{omag}, the change of action as
	\begin{eqnarray}&&
		\delta_{D}\mathcal{L}=\p_{t}\big(tE_{i}^{a}E_{i}^{a}\big)+\p_{k}\big(x^{k}E_{i}^{a}E_{i}^{a}\big)+\p_{t}\big(-\frac{1}{4}f_{ij}^{a}f_{ij}^{a}\big)+\p_{k}\big(-\frac{1}{4}f_{ij}^{a}f_{ij}^{a}\big)\\&&
		\hspace{2cm}\text{if} \quad \Delta_{1}=\Delta_{2}=1, \quad \Delta_{\xi}=2.
	\end{eqnarray}
	So the action is invariant under dilatation in four spacetime dimensions if the scaling dimensions of the temporal and spatial component of gauge fields are one, and for $\xi_{i}$ scaling dimension is two.
	
	\paragraph*{Spatial SCT:}
	Transformation of fields $a_{t}^{a}$, $a_{i}^{a}$ and $\xi_{i}^{a}$ under spatial SCT  are given by
	\bes\label{0magKi}
	\begin{eqnarray}&&
		\hspace{-1cm}\delta_{K_{l}}a_{t}^{a}=\big(2x_{l}+2x_{l}t\p_{t}+2x^{k}x_{l}\p_{k}-x_{k}x_{k}\p_{l}\big)a_{t}^{a}+2tqa_{l}^{a},\\&&
		\hspace{-1cm}\delta_{K_{l}}a_{i}^{a}=\big(2x_{l}+2x_{l}t\p_{t}+2x^{k}x_{l}\p_{k}-x_{k}x_{k}\p_{l}\big)a_{i}^{a}+2\delta_{li}x_{k}a_{k}^{a}-2\delta_{lk}x_{i}a_{k}^{a}+2tq'\delta_{li}a_{t}^{a},\\&&
		\hspace{-1cm}\delta_{K_{l}}\xi_{i}^{a}=\big(4x_{l}+2x_{l}t\p_{t}+2x^{k}x_{l}\p_{k}-x_{k}x_{k}\p_{l}\big)\xi_{i}^{a}+2\delta_{li}x_{k}\xi_{k}^{a}-2\delta_{lk}x_{i}\xi_{k}^{a}+2tq''\delta_{li}a_{t}^{a}\non\\&&\hspace{10.5cm}+2q'''tf_{il}^{a},
	\end{eqnarray}
	\ees
	using these transformations in \eqref{omag}, the action changes as
	\begin{eqnarray}&&
		\delta_{K_{l}}\mathcal{L}^{(0)}=\p_{t}(2x_{l}t\xi_{i}^{a}E_{i}^{a(0)})-\p_{t}\big(\frac{1}{2}tx_{l}f_{ij}^{a}f_{ij}^{a}\big)+\p_{k}\big(2x_{k}x_{l}\xi_{i}^{a}E_{i}^{a}\big)-\p_{k}\big(\frac{1}{2}x_{k}x_{l}f_{ij}^{a}f_{ij}^{a}\big)\non\\&&\hspace{6.7cm}-\p_{l}\big(x_{k}x_{k}\xi_{i}^{a}E_{i}^{a(0)}\big)+\p_{l}\big(\frac{1}{4}x_{k}x_{k}f_{ij}^{a}f_{ij}^{a}\big).
	\end{eqnarray}
	So the action is invariant under spatial special conformal transformation if $q=0,\,q'=1,\,q''=0,\,q'''=-1$.
	\paragraph*{Supertranslation: }
	Lastly, invariance under supertranslation ($M_{f}$), which contains Hamiltonian, temporal spacial conformal, and boost operator for different choice of $f$. Under this operator, fields transform as
	\begin{eqnarray}\label{0magMf}&&
		\hspace{-.7cm}\delta_{M_{f}}a_{t}^{a}=f(x)\p_{t}a_{t}^{a},\quad
		\delta_{M_{f}}a_{i}^{a}=f(x)\p_{t}a_{i}^{a}-a_{t}^{a}\p_{i}f(x),\quad
		\delta_{M_{f}}\xi_{i}^{a}=f(x)\p_{t}\xi_{i}^{a}-f_{ik}^{a}\p_{k}f(x),
	\end{eqnarray}
	using these in \eqref{omag}, the actions changes as
	\begin{eqnarray}
		\delta_{M_{f}}\mathcal{L}^{(0)}=\p_{t}\big(f(x)\xi_{i}^{a}E_{i}^{a(0)}-\frac{1}{4}f_{ij}^{a}f_{ij}^{a}\big)+\p_{i}\big(-\frac{1}{2}f_{ij}^{a}E_{j}^{a}\big)+\p_{j}\big(\frac{1}{2}f_{ij}^{a}E_{i}^{a}\big).
	\end{eqnarray}
	The action is invariant under $M_{f}$.
	
	Now we conclude the NLO Lagrangian or the magnetic sector action for the $\lambda=0$ case  Eq.\eqref{omag} is invariant under full infinite CCA symmetry in four spacetime dimensions.
	\subsection*{Energy-Momentum tensor} 
	Like the electric sector, the NLO Lagrangian or magnetic sector is infinite Carrollian conformal invariant in $4d$ spacetime. Now let's see what the energy-momentum tensor for the action Eq.\eqref{omag} is and see how we can improve it.
	Energy-momentum tensor of the action \eqref{omag} is given by
	\begin{eqnarray}&&
		T^{t}\,_{i}=\xi_{k}^{a}\p_{i}a_{k}^{a},\quad T^{i}\,_{t}=-x_{i}^{a}\p_{t}a_{t}^{a}-f_{ik}^{a}\p_{t}a_{k}^{a},\\&&
		T^{t}\,_{t}=\xi_{i}^{a}\p_{t}a_{k}^{a}-\mathcal{L},\quad T^{i}\,_{j}=-\xi_{i}^{a}\p_{j}a_{t}^{a}-f_{ik}^{a}\p_{j}a_{k}^{a}-\delta^{i}_{j}\mathcal{L}.
	\end{eqnarray}
	Here we also need an improved energy-momentum tensor as the electric sector. Following \cite{deBoer:2021jej},
	the improved energy-momentum tensor formula for the magnetic sector is
	\begin{equation}\label{improve em magnetic}
		T^{\mu}{_\nu}
		=
		-
		\frac{\delta\mathcal{L}}{\delta\partial_{\mu}a_{\alpha}^{a}}\partial_{\nu}a_{\alpha}^{a}
		+
		\delta^{\mu}{_\nu}\mathcal{L}
		-
		\delta^{\mu}{_t}
		\partial_{i}
		\left[
		\xi^{a}_{i}a_{\nu}^{a}
		\right]
		+
		\delta^{\mu}{_i}
		\left[
		\partial_{t}(\xi_{i}^{a}a_{\nu}^{a})
		+
		\partial_{j}(f_{ij}^{a}a_{\nu}^{a})
		\right]
		\,.
	\end{equation} 
	If we write it components wise explicitly
	\begin{eqnarray}&&
		T^{t}\,_{i}=\xi_{k}^{a}f_{ik}^{a},\quad T^{i}\,_{t}=0,\quad
		T^{t}\,_{t}=\frac{1}{4}f_{ij}^{a}f_{ij}^{a},\quad T^{i}\,_{j}=-f_{ik}^{a}f_{jk}^{a}-\delta^{i}_{j}\mathcal{L}.
	\end{eqnarray}
	Here we can see energy-momentum tensor is gauge invariant, traceless, symmetric in spatial indices, and $T^{i}_{t}=0$ as expected for Carroll theory. Using equations of motion and Carrollian Bianchi identity, we can see
	\bes
	\begin{eqnarray}\label{cem}&&
		\p_{t}T^{t}\,_{t}+\p_{i}T^{i}\,_{t}=0,\quad \text{using} \quad \eqref{0meom},\eqref{0bi1},\\&&
		\p_{t}T^{t}\,_{j}+\p_{i}T^{i}\,_{j}=0,\quad\text{using} \quad \eqref{0meom},\eqref{0bi2}.
	\end{eqnarray}
	\ees
	the energy-momentum tensor satisfies conservation equations.
	
\section{Noether charges and Charge algebra}\label{Noether charge and algebra}
All the four sectors of Carrollian Yang-Mills theory are invariant under infinite Carrollian conformal symmetry. In this section, we will study Noether's charges and charge algebra and see if there is any central extension for any commutation relation.

If we vary the Lagrangian ($L=\int d^{d-1}x \,\mathcal{L}$) on-shell on the  field space in an arbitrary direction: $\varphi \rightarrow \varphi +\delta \varphi$, we have
\bea{}\label{theta}
\delta L= \int d^{d-1}x \:\Big[\p_t {\Theta(\varphi, \p \varphi, \delta \varphi)}\Big] ~~: \text{on-shell.}
\eea
Here the expression for the $\Theta$ for all four sectors of action are
\begin{eqnarray}&&
	\delta=0:\quad
	\text{Electric Sector} \, \Theta=\delta a_{i}^{a}E_{i}^{a(0)}.\quad
	\text{Magnetic Sector}  \,\Theta=\delta a_{i}^{a}\xi_{i}^{a}.\\&&
	\delta=1:\quad
	\text{Electric Sector} \, {\Theta}=\delta a_{i}^{a}\tilde{E}_{i}^{a(0)}.\quad
	\text{Magnetic Sector}  \,{\Theta}=\delta a_{i}^{a}\xi_{i}^{a}.\label{1noethercharge}
\end{eqnarray}
Next, we consider a specific infinitesimal transformation  $\varphi \to \varphi +\delta_{\epsilon} \varphi$ off-shell. The variation $\delta_{\epsilon}$ is said to be a symmetry, if:
\bea{}\label{alpha}
\delta_{\e} L= \int d^{d-1}x \:\Big[\p_t \b  (\varphi, \p \varphi, \delta_{\e} \varphi) \Big] ~~: \text{off-shell,}
\eea
for some function $\b$ in field space.
If we compare \eqref{theta} and \eqref{alpha}, we deduce that on-shell:
\bea{}\label{inter}
\p_t Q_{\e} := \int d^{d-1}x \p_t \left( \Theta(\Phi, \p \Phi, \delta_{\epsilon} \Phi) - \b  (\Phi, \p \Phi, \delta_{\e} \Phi) \right) =0.
\eea

Noether Charge  is $Q=\int d^{d-1}x\big(\Theta-\beta$\big). Charges are listed below for different cases
\section*{Electic($\delta=0$)}
\bes
\begin{eqnarray}&&
	Q_{Boost}=\int d^{3}x \big[x_{k}\p_{t}a_{i}^{a(0)}E_{i}^{a(0)}+a_{t}^{a(0)}E_{i}^{a(0)}-\frac{x_{k}}{2}E_{i}^{a(0)}E_{i}^{a(0)}\big],\\&&
	Q_{Dilation}=\int d^{3}x \big[t\p_{t}a_{i}^{a(0)}E_{i}^{a(0)}+x^{k}\p_{k}a_{i}^{a(0)}E_{i}^{a(0)}+a_{i}^{0}E_{i}^{a(0)}-tE_{i}^{a(0)}E_{i}^{a(0)}\big],\\&&
	Q_{Spatial \,\,SCT}=\int d^{3}x \big[2x_{k}t\p_{t}a_{i}^{a(0)}E_{i}^{a(0)}+2x_{l}x_{k}\p_{l}a_{i}^{a(0)}-x_{l}x_{l}\p_{k}a_{i}^{a(0)}E_{i}^{a(0)}+2x_{k}a_{i}^{a(0)}E_{i}^{a(0)}\non\\&&
	\hspace{2.5cm}+2x_{l}a_{l}^{a(0)}E_{k}^{a(0)}-2x_{i}a_{k}^{a(0)}E_{i}^{a(0)}-tx_{k}E_{i}^{a(0)}E_{i}^{a(0)}\big],\\&&
	Q_{M_{f}}=\int d^{3}x \big[f(x)\p_{t}a_{i}^{a(0)}E_{i}^{a(0)}-a_{t}^{a(0)}\p_{i}f(x)E_{i}^{a(0)}-f(x)E_{i}^{a(0)}E_{i}^{a(0)}\big].
\end{eqnarray}
\ees
\section*{Magnetic($\delta=0$)}
\bes
\begin{eqnarray}&&
	Q_{Boost}=\int d^{3}x \big[x_{k}\p_{t}a_{i}^{a}\xi_{i}^{a}-a_{t}^{a}\xi_{i}^{a}-x_{k}\xi_{i}^{a}E_{i}^{a}+x_{k}\frac{1}{4}F_{ij}^{a}F_{ij}^{a}\big],\\&&
	Q_{Dilation}=\int d^{3}x \big[t\p_{t}a_{i}^{a}\xi_{i}^{a}+x^{k}\p_{k}a_{i}^{a}\xi_{i}^{a}+a_{i}^{a}\xi_{i}^{a}-t\xi_{i}^{a}E_{i}^{a}+\frac{1}{4}tf_{ij}^{a}f_{ij}^{a}\big],\\&&
	Q_{Spatial \,\,SCT}=\int d^{3}x \big[2x_{k}t\p_{t}a_{i}^{a}\xi_{i}^{a}+2x_{l}x_{k}\p_{l}a_{i}^{a}\xi_{i}^{a}-x_{l}x_{l}\p_{k}a_{i}^{a}\xi_{i}^{a(0)}+2x_{k}a_{i}^{a}\xi_{i}^{a}\non\\&&
	\hspace{2.5cm}+2x_{l}a_{l}^{a}\xi_{k}^{a(0)}-2x_{i}a_{k}^{a}\xi_{i}^{a}+2t\xi_{k}^{a}a_{t}^{a}-2tx_{k}E_{i}^{a}\xi_{i}^{a}-\frac{1}{2}tx_{k}f_{ij}^{a}f_{ij}^{a}\big],\\&&
	Q_{M_{f}}=\int d^{3}x \big[f(x)\p_{t}a_{i}^{a(0)}\xi_{i}^{a}+a_{t}^{a(0)}\p_{i}f(x)\xi_{i}^{a}-f(x)\xi_{i}^{a}E_{i}^{a(0)}-\frac{1}{4}f_{ij}^{a}f_{ij}^{a}\big].
\end{eqnarray}
\ees
Similarly, for the electric and the magnetic sector of the  $\delta=1$ case, we can write Noether's charge using Eq.\eqref{1noethercharge}.
Expression of charges for $\delta=1$ case are similar to $\delta=0$ case, but instead of $E_{i}^{a}$ and $f_{ij}^{a}$ we have to write in term of $\tilde{E}_{i}^{a}$ and $\tilde{f}_{ij}^{a}$ respectively.


If we check the charge algebra using these charges, there is no central extension for any commutation. In the Galilean Yang-Mills case, there is a non-trivial state-dependent central charge in the charge algebra \cite{Bagchi:2022twx}. Here we will give just one example for the electric sector, and all the other commutation relations can be realized similarly. Using the expression of $\Theta$, we can define the Poisson bracket  for the electric sector as
\begin{eqnarray}
	\Omega(\delta_{1},\delta_{2})=\delta_{1}\Theta(\delta_{2})-\delta_{2}\Theta(\delta_{1})=\delta_{1}a_{i}^{a(0)}\delta_{2}E_{i}^{a(0)}-\delta_{2}a_{i}^{a(0)}\delta_{1}E_{i}^{a(0)}.
\end{eqnarray} 
Now if we check the algebra between dilatation $(D)$ and supertranslation $(M_{f})$ using transformations of different fields in the electric sector, we can see that
\begin{eqnarray}
	\Omega(D,M_{f})=Q_{M_{h}}, \quad \text{where} \quad h=x^{k}\p_{k}f(x)-f(x).
\end{eqnarray} 
Here we can see that the commutation relation between dilatation and supertranslation is satisfied and there is no central charge. Similarly, we can realize all other commutation relations of infinite CCA for all the four sectors. A detailed discussion of the charge algebra is given in Appendix \ref{Charge Algebra app}. 
\medskip

\section{Propagator and Vertices}\label{Quantum}
We discussed the construction of Carrollian Yang-Mills actions, symmetry of all four sectors, energy-momentum tensor and its conservation, Noether charges, and charge algebra. Now we will start the Quantum aspects of the theory. Details discussion on quantum properties will be in our subsequent work; here will give all the propagators and vertices for all four sectors. 

For $\delta=1$ sector will discuss first. In this case, Lagrangian contained only kinetic terms. So there will be only propagators. After that, we will discuss the propagator, vertices, and Feynman diagram for the $\delta=0$ cases. 
\subsection{Electric Sector I}

For the $\delta=1$ case, similar to the relativistic case, we cannot calculate the propagator without adding a gauge fixing term. The full electric sector($\delta=1$) Lagrangian with a gauge fixing term is 
\begin{eqnarray}\label{1elecg}
	\mathcal{L}=\frac{1}{2}(\p_{t}a_{i}^{a(0)}-\p_{i}a_{t}^{a(0)})(\p_{t}a_{i}^{a(0)}-\p_{i}a_{t}^{a(0)})-\frac{1}{2\chi}\p_{t}a_{t}^{a}\p_{t}a_{t}^{a},
\end{eqnarray}
where $\chi$ is gauge parameter.
In order to get propagators from this kinetic part of the Lagrangian, let us first introduce Fourier transformation to momentum space
\begin{equation}\label{CYM-fourier-transform}
	\Phi^a(t,\vec{x}) = \int \frac{d\omega}{2\pi}\frac{d^3\vec{k}}{(2\pi)^3}e^{-i\omega t}e^{i\vec{k}\cdot\vec{x}}\tilde{\Phi}^a(\omega,\vec{k}),
\end{equation}
where $\Phi^a = (a_t^a, a_i^a, c^a, \bar{c}^a, \xi_{i}^{a})$, and delta functions
\begin{equation}\label{CYM-delta-functions}
	\int\frac{dt}{2\pi}e^{-i\omega t} = \delta(\omega), \quad \int \frac{d^3\vec{x}}{(2\pi)^3} e^{i\vec{k}\cdot\vec{x}} = \delta^{(3)}(\vec{k}).
\end{equation}
We also introduce the notation, $k = (\omega,\vec{k})$ and $\mathcal{A}_I^a = (a_t^a,a_i^a)$. Taking Fourier transformation and using delta functions, the action becomes
\begin{equation}\label{GYM-kin-action-4d}
	\mathcal{S} = \int \frac{d\omega d^3\vec{k}}{(2\pi)^4}\Big(\frac{1}{2}\mathcal{A}_I^a(k) d^{IJab}\mathcal{A}_J^b(-k)\Big),
\end{equation}
where
\begin{eqnarray}\label{electricmm}
	d^{IJab}(k) = i\delta^{ab}\begin{pmatrix}
		-k^{2}+\frac{\omega^{2}}{\chi} &\,\,\,\, -\omega k_{j}\\
		-\omega k_{i} &\,\,\, -\omega^{2}\delta_{ij}
	\end{pmatrix}.
\end{eqnarray}
	Then from the inverse of $d^{IJab}$, we get the propagators for the fields $\mathcal{A}_I^a$ as
	
	\begin{eqnarray}\label{pe}
		\langle \mathcal{A}_I^a\mathcal{A}_j^b\rangle=-{i\delta^{ab}}\begin{pmatrix}
			\frac{\chi}{\omega^{2}} & -\frac{k_{i}\chi}{\omega^{3}} \\
			-\frac{k_{j}\chi}{\omega^{3}} & \,\,\, \frac{k_{i}k_{j}\chi-\omega^{2}\delta_{ij}}{\omega^{4}}
		\end{pmatrix}.
	\end{eqnarray}
	where $\mathcal{A}_I^a=a_{t}^{a},a_{i}^{a}$.
			\begin{figure}[h]
				\centering
				\subfigure[Gauge field propagator $\langle \mathcal{A}^a_I\mathcal{A}^b_J\rangle$]{
					\label{fig:first}
					\includegraphics[height=1.8cm]{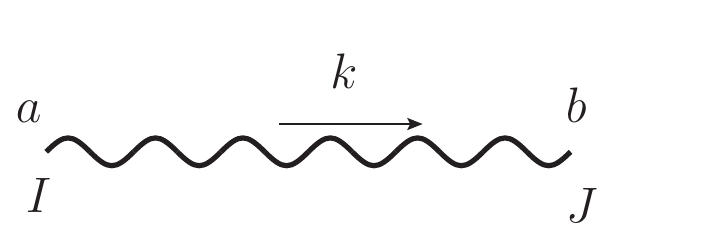}}
				\caption{ Electric  Propagator ($\delta=1$)}
			\end{figure}
			\subsection{Magnetic sector I}
			Let's consider the magnetic sector Lagrangian for $\delta=1$ before adding any gauge fixing term. The action \eqref{1magaction} explicitly in terms of gauge fields is 
			\begin{eqnarray}&&
				\mathcal{L}=
				\xi_{i}^{a}(\p_{t}a_{i}^{a(0)}-\p_{i}a_{t}^{a(0)})(\p_{t}a_{i}^{a(0)}-\p_{i}a_{t}^{a(0)})-\frac{1}{4}(\p_{i}a_{j}^{a}-\p_{j}a_{i}^{a})(\p_{i}a_{j}^{a}-\p_{j}a_{i}^{a}).
			\end{eqnarray}
			Because there is no interaction term in the above action, no vertices are possible; only propagators will be there. If we write the above action in momentum space using Eq.\eqref{CYM-fourier-transform} and Eq.\eqref{CYM-delta-functions}, we have 
			\begin{equation}\label{CYM-kin-action-4d-Magnetic}
				\mathcal{S} = \int \frac{d\omega d^3\vec{k}}{(2\pi)^4}\Big(\frac{1}{2}\mathcal{A}_I^a(k) d^{IJab}\mathcal{A}_J^b(-k)\Big),
			\end{equation}
			where
			\begin{eqnarray}
				d^{IJab}(k) = i\delta^{ab}\begin{pmatrix}
					\quad \big(0\big)_{3\times 3}
					\,\,\,\quad \big({ik_{i}}\big)_{3\times 1} \quad \quad -i\big(\omega\delta_{ij}\big)_{3\times 3}\\
					\big({ik_{i}}\big)_{1\times 3} \quad \quad 0_{1\times 1}\quad\quad \quad\quad \quad \big(0\big)_{1\times 3}\\
					-i\big(\omega\delta_{ij}\big)_{3\times 3}\quad \big(0\big)_{3\times 1} \quad \big(k^{2}\delta_{ij}-k_{i}k_{j}\big)_{3\times 3}
				\end{pmatrix}.
			\end{eqnarray}
			The determinant of this matrix is zero, so we cannot derive a propagator by doing the inverse of this matrix. We need to add a gauge fixing term, so the full Lagrangian with gauge fixing term is 
			\begin{eqnarray}\label{1magg}&&
				\hspace{0cm}\mathcal{L}=\xi_{i}^{a}(\p_{t}a_{i}^{a(0)}-\p_{i}a_{t}^{a(0)})(\p_{t}a_{i}^{a(0)}-\p_{i}a_{t}^{a(0)})-
				\frac{1}{4}(\p_{i}a_{j}^{a}-\p_{j}a_{i}^{a})(\p_{i}a_{j}^{a}-\p_{j}a_{i}^{a})\non\\&&\hspace{10.5cm}-\frac{1}{2\chi}\p_{i}a_{i}^{a}\p_{j}a_{j}^{a}.
			\end{eqnarray}
			Similar to equation\eqref{CYM-kin-action-4d-Magnetic} when we write this gauge fixed Lagrangian in momentum space the matrix $(d^{IJab}(k))$ now become
			\begin{eqnarray}\label{1magcmm}
				d^{IJab}(k) = \delta^{ab}\begin{pmatrix}
					\hspace{-1cm} \big(0\big)_{3\times 3}
					\,\,\,\quad \qquad\big({ik_{i}}\big)_{3\times 1} \quad \quad -i\big(\omega\delta_{ij}\big)_{3\times 3}\\
					\hspace{-1cm}\big({ik_{i}}\big)_{1\times 3} \quad\qquad \quad 0_{1\times 1}\quad\quad \quad\quad \quad \big(0\big)_{1\times 3}\\
					-i\big(\omega\delta_{ij}\big)_{3\times 3}\qquad\qquad \big(0\big)_{3\times 1} \quad \big(k^{2}\delta_{ij}-(1-\frac{1}{\chi})k_{i}k_{j}\big)_{3\times 3}
				\end{pmatrix}.
			\end{eqnarray}
			The determinant of this matrix is non-zero. The propagator, by doing the inverse of the above matrix is 
			\begin{eqnarray}\label{1maggpro}
				\langle \mathcal{A}_I^a\mathcal{A}_J^b\rangle=\delta^{ab}\begin{pmatrix}
					\quad \big(\frac{k^{2}\delta_{ij}-k_{i}k_{j}}{\omega^{2}}\big)_{3\times 3}
					\,\,\,\quad \big(\frac{-ik_{j}}{k^{2}}\big)_{3\times 1} \quad \quad i\big(\frac{k^{2}\delta_{ij}-k_{i}k_{j}}{\omega k^{2}}\big)_{3\times 3}\\
					\big(\frac{-ik_{i}}{k^{2}}\big)_{1\times 3} \quad \quad\quad\quad \frac{\omega^{2}\chi}{k^{4}}\quad\quad \quad\quad \big(\frac{k_{j}\omega \chi}{k^{4}}\big)_{1\times 3}\\
					i\big(\frac{k^{2}\delta_{ij}-k_{i}k_{j}}{\omega k^{2}}\big)_{3\times 3}\quad \big(\frac{k_{j}\omega \chi}{k^{4}}\big)_{3\times 1} \quad\quad\quad \big(\frac{-k_{i}k_{j}\chi}{k^{4}}\big)_{3\times 3}
				\end{pmatrix}.
			\end{eqnarray}
			where $\mathcal{A}_I^a=\xi_{i}^{a},a_{t}^{a},a_{i}^{a}$.
			\begin{figure}[h]
				\centering
				\subfigure[Gauge field propagator $\langle \mathcal{A}^a_I\mathcal{A}^b_J\rangle$]{
					\label{fig:first}
					\includegraphics[height=1.8cm]{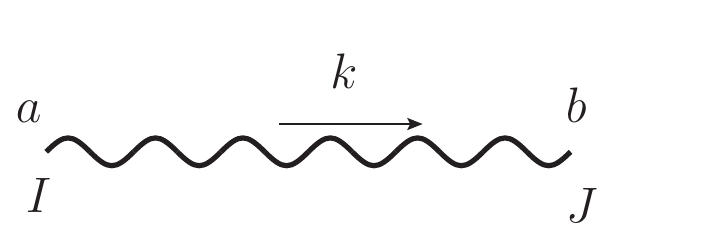}}
				\caption{ Magnetic  Propagator ($\delta=1$)}
			\end{figure}
			
			This section considered Lagrangian for the $\delta=1$ case for the electric and magnetic sectors. These Lagrangians only contain kinetic terms, so there are only propagators, not vertices.

			\subsection{Electric sector II}\label{epro}
			Now focus on the $\delta=0$ cases for propagator and vertices containing non-abelian or self-interaction terms. The full Lagrangian for the $\delta=0$ electric sector with gauge fixing term and ghost term is 
			\begin{eqnarray}\label{0elecgg}
				\mathcal{L}=\frac{1}{2}E_{i}^{a(0)}E_{i}^{a(0)}-\frac{1}{2\chi}\p_{t}a_{t}^{a}\p_{t}a_{t}^{a}++\p_{t}\bar{c}^{a}D_{t}c^{a}.
			\end{eqnarray}
			The kinetic part of the above Lagrangian is
			\begin{eqnarray}
				\mathcal{L}_{kin}=\frac{1}{2}(\p_{t}a_{i}^{a(0)}-\p_{i}a_{t}^{a(0)})(\p_{t}a_{i}^{a(0)}-\p_{i}a_{t}^{a(0)})-\frac{1}{2\chi}\p_{t}a_{t}^{a}\p_{t}a_{t}^{a}+\p_{t}\bar{c}^{a}\p_{t}c^{a}.
			\end{eqnarray} 
			In momentum space using equations \eqref{CYM-fourier-transform} and \eqref{CYM-delta-functions} we can write
			\begin{equation}\label{GYM-kin-action-4d}
				\mathcal{S}_{kin} = \int \frac{d\omega d^3\vec{k}}{(2\pi)^4}\Big(\frac{1}{2}\mathcal{A}_I^a(k) d^{IJab}\mathcal{A}_J^b(-k) +\bar{c}^a(k)\big(-\omega^{2}\big)c^a(-k) \Big),
			\end{equation}
			The kinetic part of gauge fields and the gauge fixing term is the same as the Eq.\eqref{1elecg}. So the expression of $d^{IJab}$ is same as  \eqref{electricmm}. The inverse of this matrix is the propagator for gauge fields given in equation \eqref{pe}. And the inverse of the coefficient of $\bar{c}c$ in \eqref{GYM-kin-action-4d} gives the propagator for ghost fields as
			\begin{eqnarray}
				\langle \bar{c}^{a}(k)c^{b}(-k)\rangle =\frac{i\delta^{ab}}{\omega^{2}}.
			\end{eqnarray}
			In a compact form the propagators and Feynman diagrams of the Lagrangian are 
			\bes
			\begin{align}\label{propagatorelect}
				\begin{split}
					&\raisebox{-0.3\height}{\includegraphics[width=2.1in]{f1.pdf}}
					\equiv -i\delta^{ab}\begin{pmatrix}
						\frac{\chi}{\omega^{2}} & -\frac{k_{i}\chi}{\omega^{3}} \\
						-\frac{k_{j}\chi}{\omega^{3}} & \,\,\, \frac{k_{i}k_{j}\chi-\omega^{2}\delta_{ij}}{\omega^{4}}
					\end{pmatrix}~,
					\\
					&\includegraphics[width=2.1in]{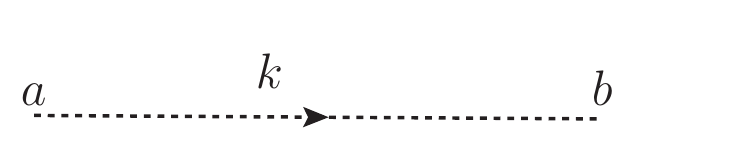}
					\equiv
					\langle \bar{c}^{a}(k)c^{b}(-k)\rangle =\frac{i\delta^{ab}}{\omega^{2}}.
				\end{split}.
			\end{align}
			\ees
			Interaction terms of the Lagrangian (Eq.\eqref{0elecgg}) are 
			\begin{eqnarray}
				\mathcal{L}_{int}=2gf^{abc}a_{t}^{b}a_{i}^{c}\p_{t}a_{i}^{c}-2gf^{abc}a_{t}^{b}a_{i}^{c}\p_{i}a_{t}^{a}+g^{2}f^{abc}f^{ade}a_{t}^{b}a_{i}^{c}a_{t}^{d}a_{i}^{e}-gf^{abc}a_{t}^{a}\p_{t}\bar{c}^{c}c^{b}.
			\end{eqnarray}
			By transforming to momentum space\eqref{CYM-fourier-transform} and using the delta functions \eqref{CYM-delta-functions} definition, we can write the three field interaction terms  as
			\begin{eqnarray}
				\mathcal{S}^{(3)}_{int} &=& \int \frac{1}{(2\pi)^{12}}{\displaystyle{\prod_{i=1}^{3}d\omega_i d^3\vec{k}_i}}\,(2\pi)^4\delta(\omega_1 + \omega_2 + \omega_3) \delta^{(3)}(\vec{k}_1 + \vec{k}_2 + \vec{k}_3) g f^{abc} \times \nonumber \\
				&& \hspace{1cm} \Big[(\omega_1 - \omega_2)a_{t}^{b}(k_{1})a_{i}^{c}(k_{2})a_{i}^{a}(k_{3}) + i\delta^{ij}(k_1 -k_2)_i a_{t}^{b}(k_{1})a_t^c(k_3)a_j^a(k_2)\non\\&&\hspace{7.4cm}-i\omega_2 \phi^a(k_1)\bar{c}^b(k_2)c^c(k_3)\Big],
			\end{eqnarray}
			where ${\displaystyle \prod_{i=1}^{n}d\omega_i d^3\vec{k}_i}=d\omega_1 d^3\vec{k}_1...d\omega_n d^3\vec{k}_n$. From this expression we can write the $3$-point vertices as
			\begin{eqnarray}
				&& V_{3\,a_t a_{i} a_{i}}^{abc} = -g f^{abc}(\omega_1 - \omega_2),\, V_{3\, a_t a_i a_{i}}^{abc\,i} = -g f^{abc}(k_1 - k_2)^i,\,V_{3\,a_{t} \bar{c} c}^{abc} = g f^{abc}\omega_2,
			\end{eqnarray}
			Similarly, transforming the four field interaction terms of $\mathcal{S}_{int}$ to momentum space, we get
			\begin{eqnarray}
				\mathcal{S}^{(4)}_{int} &=& \int \frac{1}{(2\pi)^{16}}{\displaystyle{\prod_{i=1}^{4}d\omega_i d^3\vec{k}_i}}\,(2\pi)^4\delta(\omega_1 + \omega_2 + \omega_3 + \omega_4) \delta^{(3)}(\vec{k}_1 + \vec{k}_2 + \vec{k}_3 + \vec{k}_4) \times \nonumber \\
				&& \hspace{5.3cm}\quad g^2 \Big[f^{abc}f^{ade}a_{t}^{b}(k_1)a_{i}^{c}(k_2)a_{t}^{d}(k_3)a_{i}^{e}(k_4) \Big],
			\end{eqnarray}
			from which we can read of the $4$-point vertices 
			\begin{eqnarray}
				&& V_{4\, a_t a_{i} a_t a_j}^{bcde} = -2ig^2\delta_{ij}\big(f^{abc}f^{ade} + f^{abe}f^{adc}\big).
			\end{eqnarray}
			\begin{figure}[h]\label{Electric Sector Feynman Diagram}
				\centering
				\subfigure[][$V_{3\, a_{t}a_{i}a_{j}}^{abc}$]{
					\label{fig:ex3-a}
					\includegraphics[height=3.3cm]{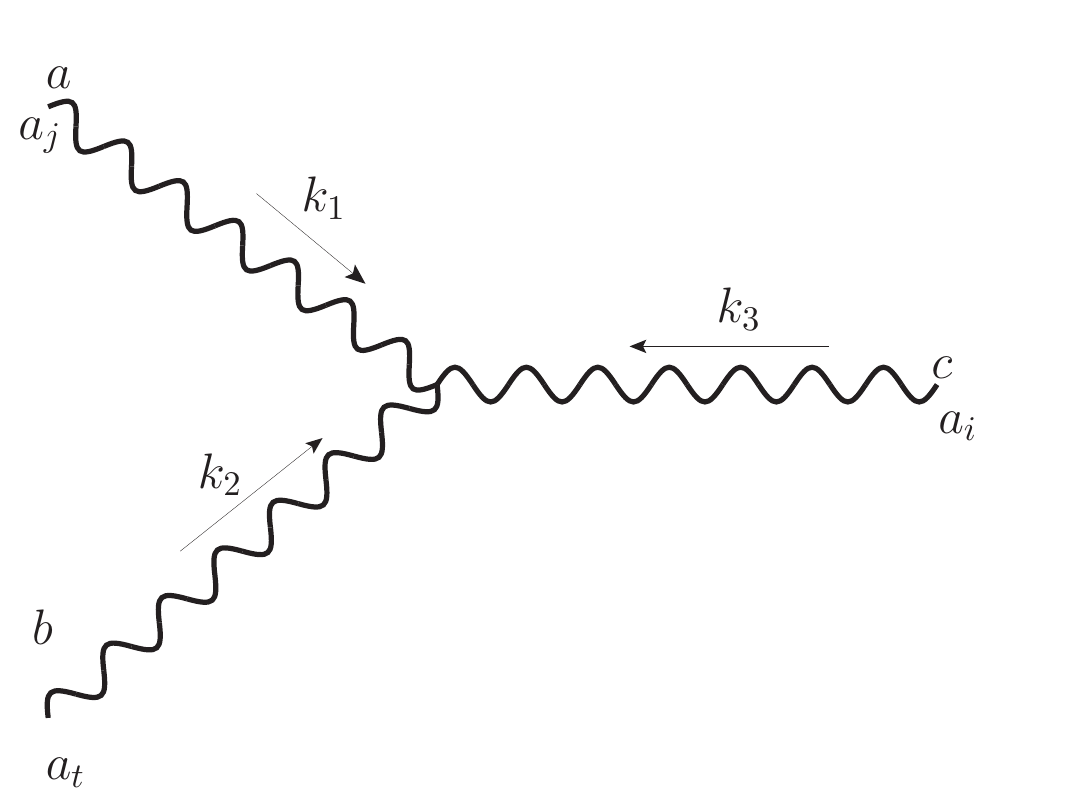}}
				\hspace{8pt}
				\subfigure[][$V_{3\, a_{t} a_t a_i}^{abc\,i}$]{
					\label{fig:ex3-b}
					\includegraphics[height=3.3cm]{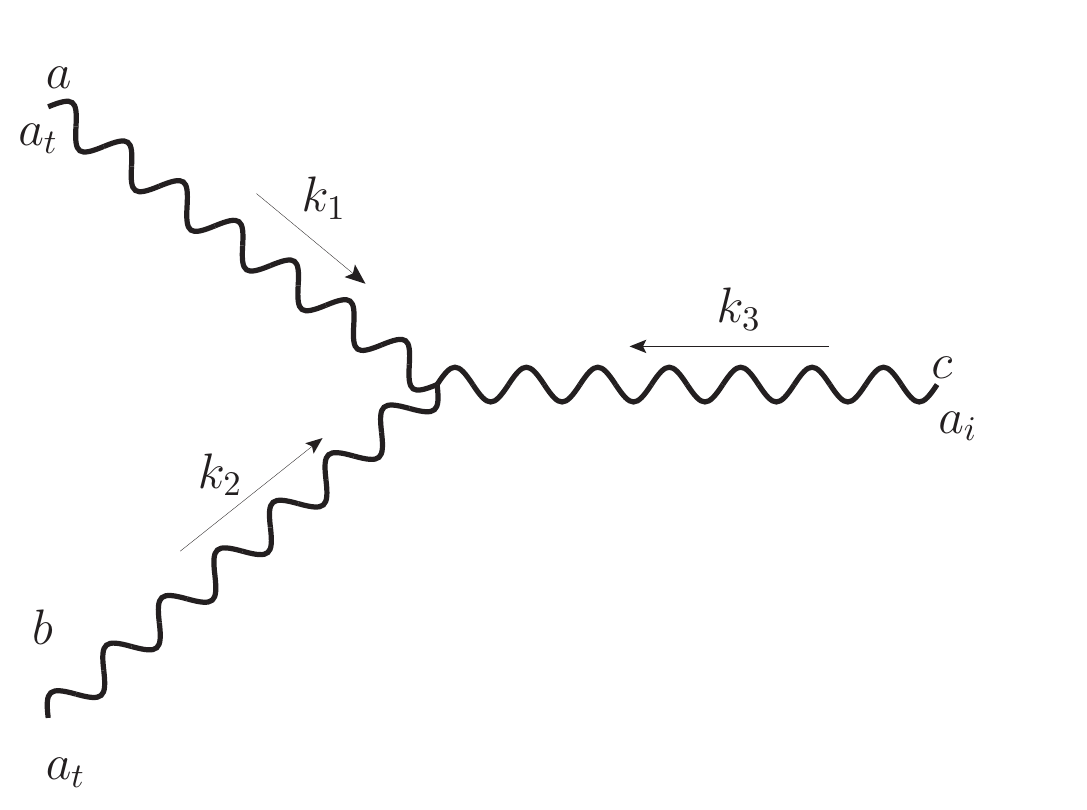}}
				\subfigure[][$V_{3\,a_{t}a_{t} a_i a_j}^{bcde\, ij}$]{
					\label{fig:ex3-c}
					\includegraphics[height=3.3cm]{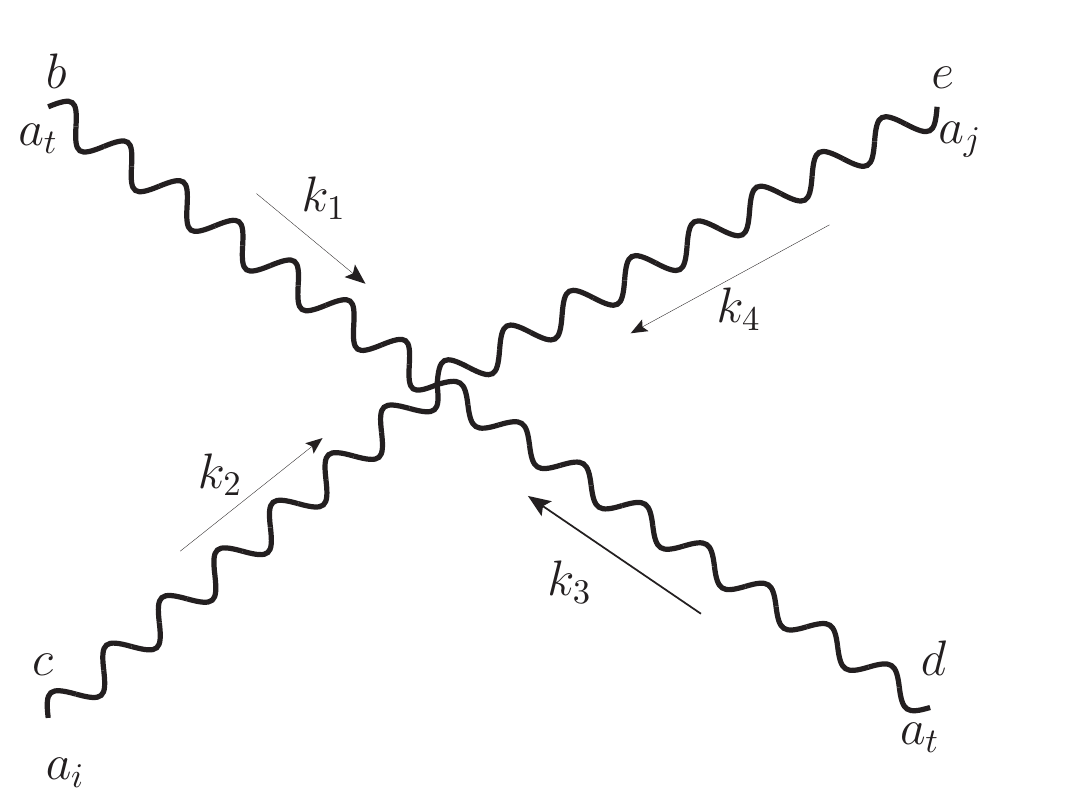}}\\
				\subfigure[][$V_{3\,a_t \bar{c} c}^{abc}$]{
					\label{fig:ex3-c}
					\includegraphics[height=3.3cm]{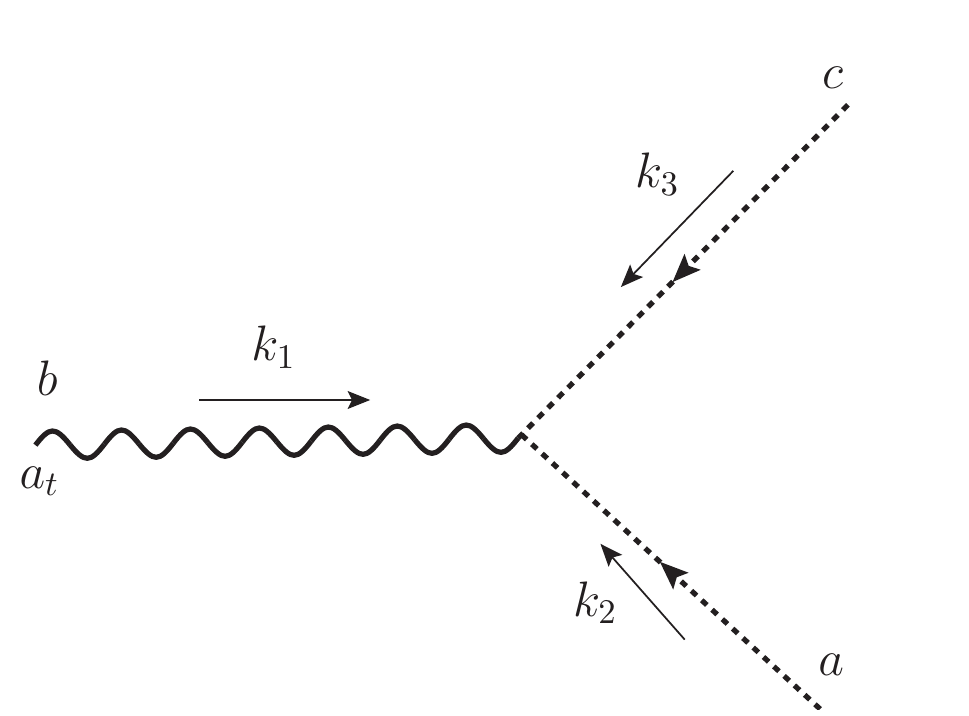}}
				\caption[Electric Sector Feynman Diagrams.]{Electric Sector Feynman Diagrams}
			\end{figure}

			\subsection{Magnetic sector II}
			
			The full magnetic sector Lagrangian for the $\delta=0$ case with gauge fixing term and ghost term is
			\begin{eqnarray}\label{1maggg}
				\mathcal{L}=\xi_{i}^{a}E_{i}^{a}-\frac{1}{4}f_{ij}^{a}f_{ij}^{a}-\frac{1}{2\chi}\p_{i}a_{i}^{a}\p_{j}a_{j}^{a}-\p_{i}\bar{c}^{a}D_{i}c^{a}.
			\end{eqnarray}
			The kinetic part of the Lagrangian is
			\begin{eqnarray}&&
				\hspace{-1cm}\mathcal{L}_{kin}=\xi_{i}^{a}(\p_{t}a_{i}^{a(0)}-\p_{i}a_{t}^{a(0)})-
				\frac{1}{4}(\p_{i}a_{j}^{a}-\p_{j}a_{i}^{a})(\p_{i}a_{j}^{a}-\p_{j}a_{i}^{a})-\frac{1}{\chi}\p_{i}a_{i}^{a}\p_{j}a_{j}^{a}-\p_{i}\bar{c}^{a}\p_{i}c^{a}.
			\end{eqnarray}
			Using equations \eqref{CYM-fourier-transform} and \eqref{CYM-delta-functions} we can write the above Lagrangian in momentum space as 
			\begin{equation}\label{GYM-kin-action-4d}
				\mathcal{S}_{kin} = \int \frac{d\omega d^3\vec{k}}{(2\pi)^4}\Big(\frac{1}{2}\mathcal{A}_I^a(k) d^{IJab}\mathcal{A}_J^b(-k) +\bar{c}^a(k)\big(-\vec{k}^{2}\big)c^a(-k) \Big),
			\end{equation}
			where
			$\mathcal{A}_{\mu}^{a}=\xi_{i}^{a},a_{t}^{a},a_{i}^{a}$.
			
			The kinetic terms of gauge fields of this magnetic sector is the same as the Eq.\eqref{1magg}, so the matrix $d^{IJab}$ is same as Eq.\eqref{1magcmm}, so as the propagator Eq.\eqref{1maggpro}. The inverse of the coefficient of $\bar{c}c$  gives the propagator for the ghost fields as
			\begin{eqnarray}
				\langle \bar{c}^{a}(k)c^{b}(-k)\rangle =\frac{i\delta^{ab}}{\vec{k}^{2}}.
			\end{eqnarray}
			In a compact form the propagators and Feynman diagram of the Lagrangian are 
			\begin{align}\label{propagatormagnetic}
				\begin{split}
					&\raisebox{-0.3\height}{\includegraphics[width=2.1in]{f1.pdf}}
					\equiv \delta^{ab}\begin{pmatrix}
						\quad \big(\frac{k^{2}\delta_{ij}-k_{i}k_{j}}{\omega^{2}}\big)_{3\times 3}
						\,\,\,\quad \big(\frac{-ik_{j}}{k^{2}}\big)_{3\times 1} \quad \quad i\big(\frac{k^{2}\delta_{ij}-k_{i}k_{j}}{\omega k^{2}}\big)_{3\times 3}\\
						\big(\frac{-ik_{i}}{k^{2}}\big)_{1\times 3} \quad \quad\quad\quad \frac{\omega^{2}\chi}{k^{4}}\quad\quad \quad\quad \big(\frac{k_{j}\omega \chi}{k^{4}}\big)_{1\times 3}\\
						i\big(\frac{k^{2}\delta_{ij}-k_{i}k_{j}}{\omega k^{2}}\big)_{3\times 3}\quad \big(\frac{k_{j}\omega \chi}{k^{4}}\big)_{3\times 1} \quad\quad\quad \big(\frac{-k_{i}k_{j}\chi}{k^{4}}\big)_{3\times 3}
					\end{pmatrix}~,
					\\
					&\includegraphics[width=2.1in]{f2.pdf}
					\equiv
					\langle \bar{c}^{a}(k)c^{b}(-k)\rangle =\frac{i\delta^{ab}}{\vec{k}^{2}}.
				\end{split}
			\end{align}
			Interaction terms of the Lagrangian are
			\begin{eqnarray}
				\mathcal{L}_{int}=gf^{abc}a_{t}^{b}a_{i}^{c}\xi_{i}^{a}-gf^{abc}a_{i}^{b}a_{j}^{c}\p_{i}a_{j}^{a}-\frac{1}{4}g^{2}f^{abc}f^{ade}a_{i}^{b}a_{j}^{c}a_{i}^{d}a_{j}^{e}-gf^{abc}a_{i}^{b}\p_{i}\bar{c}^{a}c^{c}.
			\end{eqnarray}
			All the three point interactions in momentum space using \eqref{CYM-fourier-transform} and \eqref{CYM-delta-functions} are
			\begin{eqnarray}
				\mathcal{S}^{(3)}_{int} &=& \int \frac{1}{(2\pi)^{12}}{\displaystyle{\prod_{i=1}^{3}d\omega_i d^3\vec{k}_i}}\,(2\pi)^4\delta(\omega_1 + \omega_2 + \omega_3) \delta^{(3)}(\vec{k}_1 + \vec{k}_2 + \vec{k}_3) g f^{abc} \times\non\\&& \big[i\delta^{ij} \xi_{i}^a(k_1)a_t^b(k_2)a_j^c(k_3)++ \frac{i}{6}\big((k_1 - k_2)_i \delta^{il}\delta^{jk} + (k_2 - k_3)_i \delta^{ij}\delta^{lk} + (k_3 - k_1)_i \delta^{ik}\delta^{jl} \big)\times \non\\&&\hspace{5.3cm}a_j^a(k_1)a_k^b(k_2)a_l^c(k_3)+ i\delta^{ij}k_{2j}a_i^a(k_1)\bar{c}^b(k_2)c^c(k_3)\big],
			\end{eqnarray}
			from where we can write the $3$-point vertices as
			\begin{eqnarray}
				&& V_{3\,\xi_{i} a_t a_i}^{abc} = -g f^{abc},\quad V_{3\,a_i \bar{c} c}^{abc} = - g f^{abc} k_2^i.\nonumber\\
				&& V_{3\,a_i a_j a_k}^{abc\, ijk} = -g f^{abc}\big( (k_1 - k_2)^k\delta^{ij} + (k_2 - k_3)^i\delta^{jk} + (k_3 - k_1)^j\delta^{ik} \big).
			\end{eqnarray}
			The four point interaction terms in momentum space are 
			\begin{eqnarray}
				\mathcal{S}^{(4)}_{int} &=& \int \frac{1}{(2\pi)^{16}}{\displaystyle{\prod_{i=1}^{4}d\omega_i d^3\vec{k}_i}}\,(2\pi)^4\delta(\omega_1 + \omega_2 + \omega_3 + \omega_4) \delta^{(3)}(\vec{k}_1 + \vec{k}_2 + \vec{k}_3 + \vec{k}_4) \times \nonumber \\
				&& \quad g^2 \Big[-\frac{1}{24}\big( f^{abc}f^{ade}(\delta^{ik}\delta^{jl} - \delta^{il}\delta^{jk}) + f^{abd}f^{ace}(\delta^{ij}\delta^{kl} - \delta^{il}\delta^{jk}) \nonumber \\
				&&\qquad  + f^{abe}f^{acd}(\delta^{ij}\delta^{kl} - \delta^{ik}\delta^{jl}) \big)a_i^b(k_1)a_j^c(k_2)a_k^d(k_3)a_l^e(k_4) \Big],
			\end{eqnarray}
			from which we can read of the $4$-point vertices as
			\begin{eqnarray}
				&& V_{4\,a_i a_j a_k a_l}^{bcde\,ijkl} = -ig^2\Big( f^{abc}f^{ade}(\delta^{ik}\delta^{jl} - \delta^{il}\delta^{jk}) + f^{abd}f^{ace}(\delta^{ij}\delta^{kl} - \delta^{il}\delta^{jk}) \nonumber \\
				&& \hspace{35mm} + f^{abe}f^{acd}(\delta^{ij}\delta^{kl} - \delta^{ik}\delta^{jl}) \Big).
			\end{eqnarray}
			\begin{figure}[h]
				\centering
				\subfigure[][$V_{3\,\xi_{i} a_{t}a_{j}}^{abc}$]{
					\label{fig:ex3-c}
					\includegraphics[height=3.3cm]{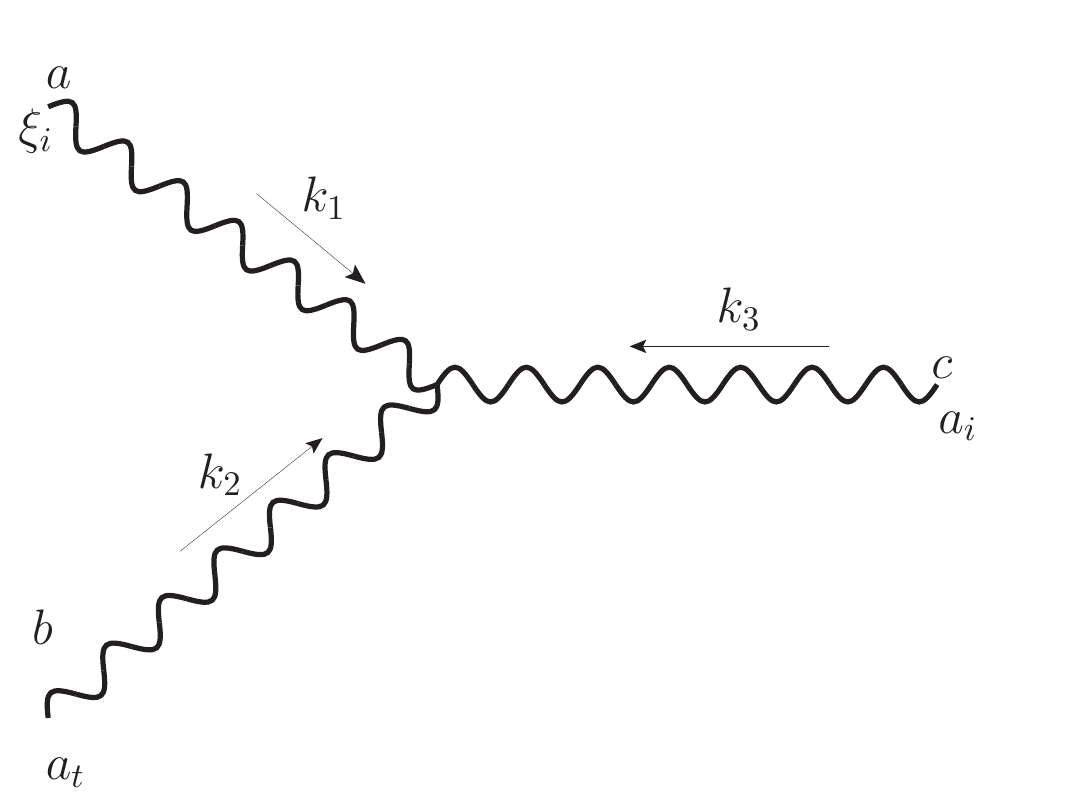}}
				\subfigure[][$V_{3\,a_{i}a_{j}a_{k}}^{abc}$]{
					\label{fig:ex3-c}
					\includegraphics[height=3.3cm]{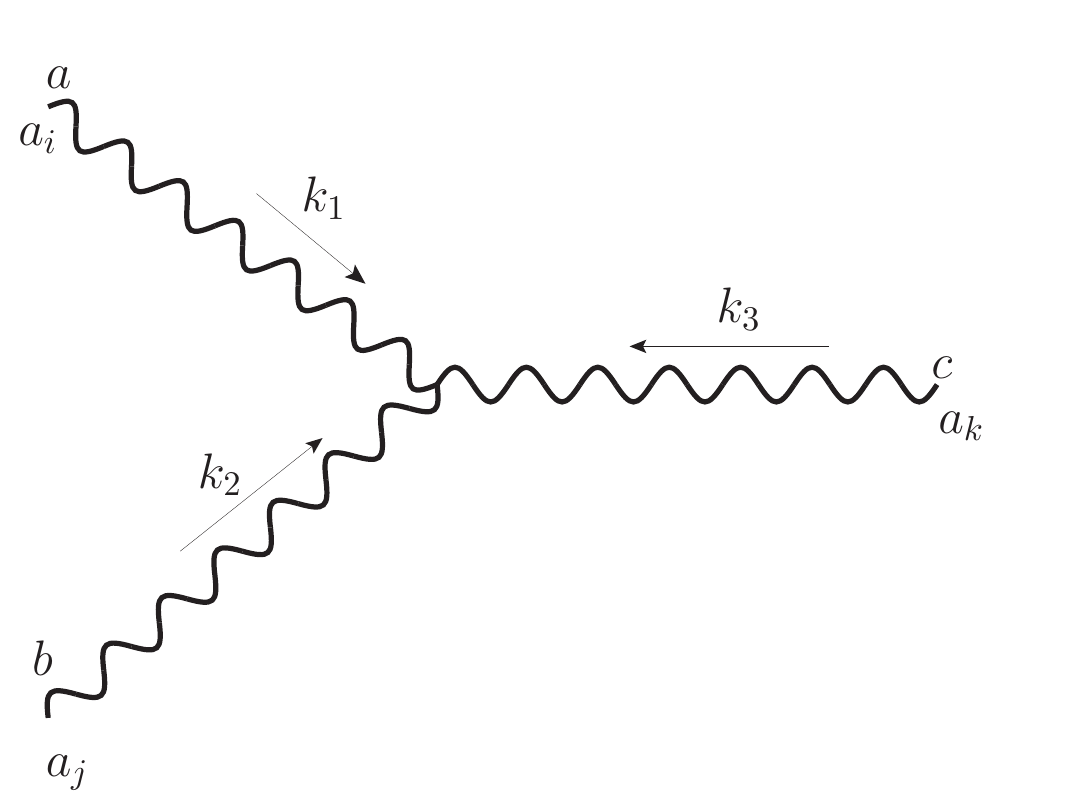}}
				\subfigure[][$V_{4\,a_i a_j a_k a_l}^{bcde\,ijkl}$]{
					\label{fig:ex3-c}
					\includegraphics[height=3.3cm]{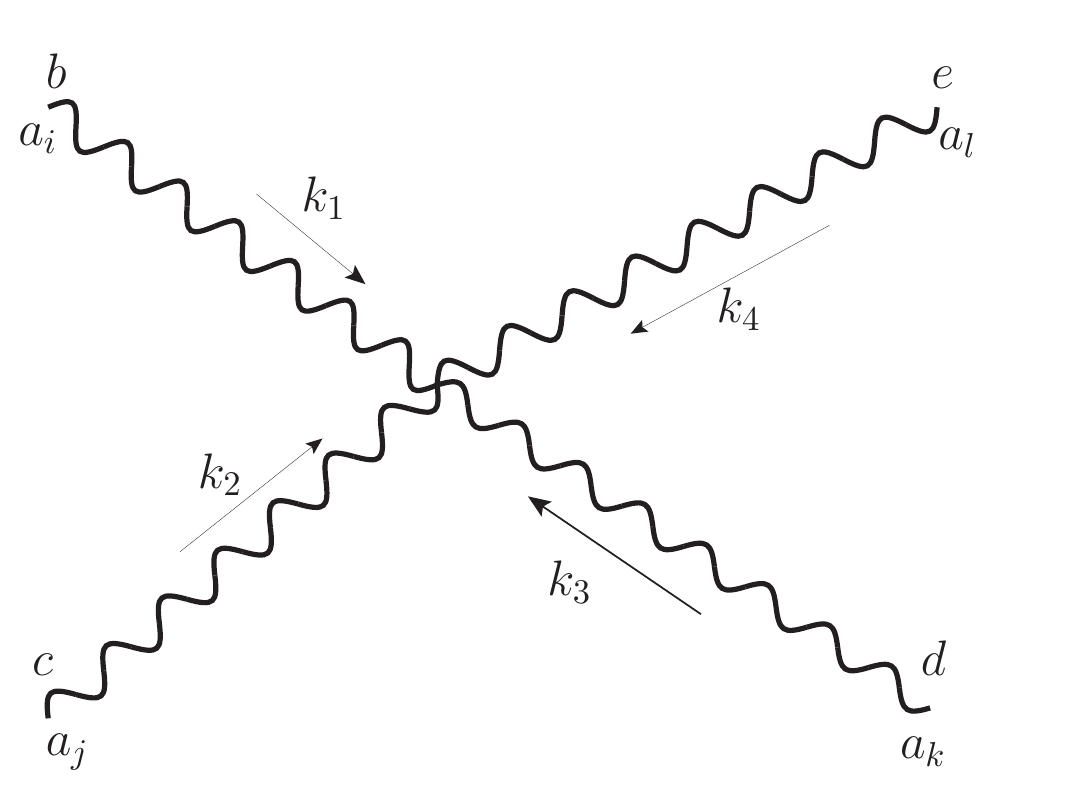}}
				\\
				\subfigure[][$V_{4\,a_i \bar{c}c}^{abc}$]{
					\label{fig:ex3-c}
					\includegraphics[height=3.3cm]{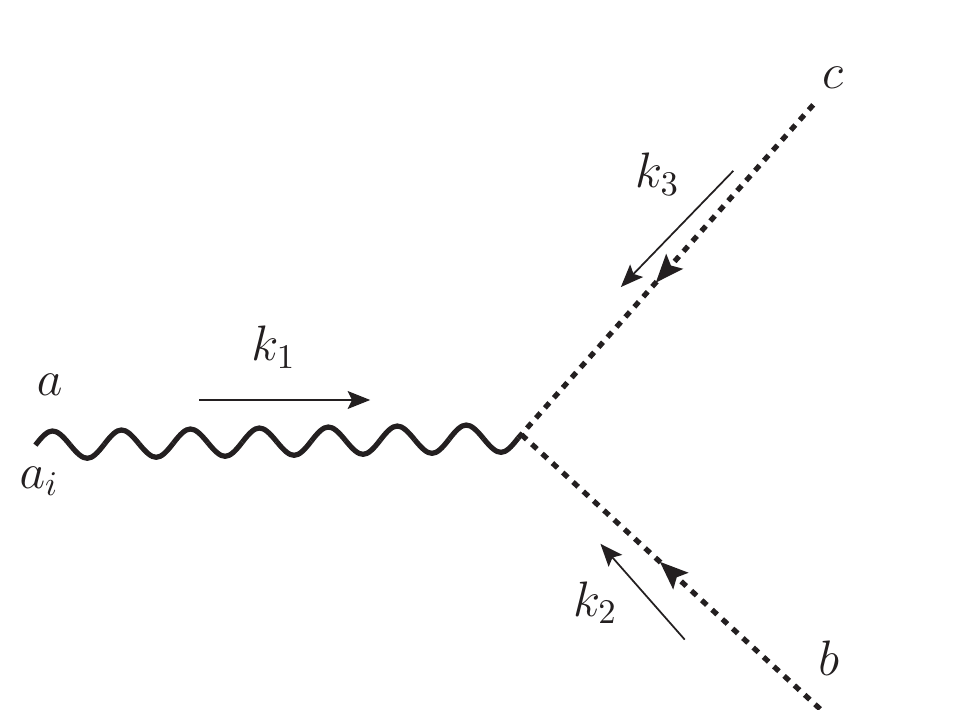}}
				\caption[Magnetic Sector Feynman Diagram.]{Magnetic Sector Feynman Diagrams}
				\label{fig:ex3}
			\end{figure}
			We will study the quantum properties of the non-trivial sectors of CYM in detail in our subsequent work using the Feynman rules listed above. After that, we will add the matter field to Carrollian Yang-Mills and construct a QCD-like structure in the Carrollian theory. 
			\subsection{Propagators in position space}
			In this section, we will  see propagator of gauge fields in position.
			In momentum space, the electric and magnetic sectors' propagators are \eqref{pe} and \eqref{1maggpro}. All the correlation function of the electric and magnetic sector in position space is listed below    
			\subsubsection*{Electric}
			
			\begin{eqnarray}&&
				\hspace{-1cm}	G_{tt}^{ab}(x-y)=\int \int \frac{d\omega d^{3}\vec{k}}{(2\pi)^{4}}\langle a_{t}^{a}a_{t}^{b}\rangle e^{-i\omega t}e^{ik^{i}x_{i}}=\delta^{ab}\chi \, 2\pi t\delta^{3}(\vec{x}),\\&&
				\hspace{-1cm}	G_{ij}^{ab}(x-y)=\int \int \frac{d\omega d^{3}\vec{k}}{(2\pi)^{4}}\langle a_{i}^{a}a_{j}^{b}\rangle e^{-i\omega t}e^{ik^{i}x_{i}}=\delta^{ab}\big(\chi\, \frac{2}{3}\pi t^{3}\p_{i}\p_{j}\delta^{3}(\vec{x})-\delta_{ij}2\pi t\delta^{3}(\vec{x})\big),\\&&
				\hspace{-1cm}	G_{ti}^{ab}(x-y)=\int \int \frac{d\omega d^{3}\vec{k}}{(2\pi)^{4}}\langle a_{t}^{a}a_{i}^{b}\rangle e^{-i\omega t}e^{ik^{i}x_{i}}=\delta^{ab}\chi \, \pi t^{2} \p_{i}\delta^{3}(\vec{x}).
			\end{eqnarray}
			We can see the propagator is of the form of $\delta(x)$ with some time function. This means there is no propagation in space, only propagation in time. This is the behaviour expected from electric versions of theories and has been observed e.g. in the theory of scalars and $U(1)$ gauge fields earlier. 
			
			\subsubsection*{Magnetic}
			
			\begin{eqnarray}&&
				\hspace{-.5cm}G_{tt}^{ab}(x-y)=\int \int \frac{d\omega d^{3}\vec{k}}{(2\pi)^{4}}\langle a_{t}^{a}a_{t}^{b}\rangle e^{-i\omega t}e^{ik^{i}x_{i}}=
				=-\delta^{ab}\p_{t}\p_{t}\delta(t)4\pi r\Lambda,\\&&
				\hspace{-.5cm}G_{ti}^{ab}(x-y)=\int \int \frac{d\omega d^{3}\vec{k}}{(2\pi)^{4}}\langle a_{t}^{a}a_{i}^{b}\rangle e^{-i\omega t}e^{ik^{i}x_{i}}=
				-\delta^{ab}\p_{t}\delta(t)4\pi \frac{x_{i}}{r}\Lambda,\\&&
				\hspace{-.5cm}G_{ij}^{ab}(x-y)=\int \int \frac{d\omega d^{3}\vec{k}}{(2\pi)^{4}}\langle a_{i}^{a}a_{j}^{b}\rangle e^{-i\omega t}e^{ik^{i}x_{i}}=4\pi\delta^{ab}\delta(t)\{\frac{\delta_{ij}}{r}-\frac{x_{i}x_{j}}{r^{3}}\},\\&&
				\hspace{-.5cm}{G_{ti}^{ab}(x-y)}_{\xi}=\int \int \frac{d\omega d^{3}\vec{k}}{(2\pi)^{4}}\langle \xi_{i}^{a}a_{t}^{b}\rangle e^{-i\omega t}e^{ik^{i}x_{i}}=
				-2\pi^{2}\delta^{ab}\delta(t)\frac{x_{i}}{r^{3}},\\&&
				\hspace{-.5cm}{G_{ij}^{ab}(x-y)}_{\xi}=\int \int \frac{d\omega d^{3}\vec{k}}{(2\pi)^{4}}\langle \xi_{i}^{a}a_{j}^{b}\rangle e^{-i\omega t}e^{ik^{i}x_{i}}=-\frac{\pi}{2}\,\delta_{ij}\,\delta^{3}(\vec{x})-2\pi^{3}\{\frac{\delta_{ij}}{r^{3}}-\frac{3x_{i}x_{j}}{r^{5}}\},\\&&
				\hspace{-.5cm}{G_{ij}^{ab}(x-y)}_{\xi\xi}=\int \int \frac{d\omega d^{3}\vec{k}}{(2\pi)^{4}}\langle \xi_{i}^{a}\xi_{j}^{b}\rangle e^{-i\omega t}e^{ik^{i}x_{i}}=\delta^{ab}2\pi t\big[\delta_{ij}\p^{2}\delta^{3}(\vec{x})-\p_{i}\p_{j}\delta^{3}(\vec{x})\big].
			\end{eqnarray}
			In the first two propagators, $\Lambda=\int_{0}^{\infty} \frac{sin\theta}{\theta^{3}}d\theta$, this is a divergent integration. To regularize it
			\begin{eqnarray}
				\Lambda=\int_{0}^{\infty} \frac{sin\theta}{\theta^{3}}d\theta=\int_{0}^{\infty} \frac{1}{\theta^{2}}d\theta-\frac{\pi}{4}=\lim_{\e\rightarrow 0}\bigg(\int_{\e}^{\infty} \frac{1}{\theta^{2}}d\theta-\frac{\pi}{4}\bigg)=\lim_{\e\rightarrow 0}\bigg(\frac{1}{\e}-\frac{\pi}{4}\bigg)
			\end{eqnarray}
			Propagators in position space have delta functions in all of the above cases, but some delta functions are of time, and some are in spatial coordinates. This finding is somewhat surprising as one does not expect to find a mixture of spatial and temporal delta functions in the magnetic sector, only temporal delta functions. This is a pointer perhaps that all Carrollian magnetic theories would not be reducible to lower dimensional Euclidean CFTs as was shown in the scalar case in \cite{Baiguera:2022lsw}. This point requires further investigation.

				\section{Conclusions and Discussions}\label{CF}
				In this paper, we have analyzed the Carrollian limit of the Yang-Mills theory systematically, and obtained electric and magnetic sectors with one subsector of each of the electric and magnetic sectors having non-abelian or self-interaction terms while the other subsector having copies of the Carrollian abelian theory. The Carrollian abelian theory found here is consistent with that discussed in \cite{deBoer:2021jej}. This is a first action formulation for the Carrollian Yang-Mills theory. We have obtained the Carrollian Yang-Mills actions by taking the a small $c$-expansion of the Poincar\'e invariant Yang-Mills action, where we observed that different values of the parameter $\delta$, used in the small $c$-expansion of the gauge fields, lead to different sectors for the Carrollian Yang-Mills theory. In particular, for $\delta=0$, we get two non-trivial Carrollian Yang-Mills theories, and for any non-zero value of $\delta$ (that we have taken to be $\delta=1$ for simplicity), we get copies of the Carrollian abelian theory. In $4$-dimensions, all these four sectors are found to be invariant under infinite CCA. The energy-momentum tensor for all the four sectors were calculated and were found to be conserved using the equations of motion and the Bianchi identities. We have also calculated Noether charges for all four sectors, and found that there are no central extensions in the algebra of the charges. This is unlike the Galilean Yang-Mills theories in \cite{Bagchi:2022twx}, where we have seen that there is a state-dependent central extension in the algebra of the charges. Finally, we listed all the Feynman rules to understand the quantum properties of the Carrollian Yang-Mills theory, with a detailed analysis kept for a future work. Further, we also calculated the propagators in position space, and from these we explicitly saw the ultra-local behavior of the Carrollian theory.
				
				\medskip
				
				There are a number of immediate directions for future work. The construction of different Carrollian Yang-Mills is our first step toward our goal of understanding the full quantum properties of Carrollian Yang-Mills theory. In our subsequent work, we will study the quantum structure of CYM theory, free and with matter fields, and investigate the different types of actions we have found in this work. In \cite{Bagchi:2022owq}, authors studied the algebraic structure of Carrollian supersymmetric theory, and in the near future, we want to construct the Carrollian version of ${\cal N}=4$ Super Yang-Mills theory and understand its role in flat space holography. The different actions we have found in this paper would be a starting point for the supersymmetrization of CYM theory. 

				%
				
				\medskip

				%
				\section*{Acknowledgments}
				We would like to first thank Arjun Bagchi for fruitful discussions, necessary suggestions, and valuable comments on the manuscript. We would also like to thank Nilay Kundu, Kedar Kolekar, and Sudipta Dutta for fruitful discussions.
				
				\appendix
				\section{Rotation and Boost invariance}\label{app:rotation-boost}
				\subsection*{Electric($\delta$=0)}
				\textbf{Rotation}:-
				Under rotation fields transform as
				\begin{eqnarray}&&
					\delta_{M_{ij}}a_{t}^{a(0)}=(x_{i}\p_{j}-x_{j}\p_{i})a_{t}^{a(0)}\\&&
					\delta_{M_{ij}}a_{k}^{a(0)}=(x_{i}\p_{j}-x_{j}\p_{i})a_{k}^{a(0)}+(\delta_{ik}a_{j}^{a(0)}-\delta_{jk}a_{i}^{a(0)})
				\end{eqnarray}
				%
				%
				The action \eqref{Electric0} changes under these transformations as
				\begin{eqnarray}
					\delta_{M_{ij}}\mathcal{L}^{(0)}=\p_{j}\big(x_{i}E_{k}^{a(0)}E_{k}^{a(0)}\big)-\p_{i}\big(x_{j}E_{k}^{a(0)}E_{k}^{a(0)}\big)
				\end{eqnarray}
				The action is rotation invariant.\\
				\textbf{Boost}:- 
				Fields transform as 
				\begin{eqnarray}&&
					\delta_{B_{i}}a_{t}^{a(0)}=x_{i}\p_{t}a_{t}^{a(0)}+q_{1}a_{i}^{(0)a}\\&&
					\delta_{B_{i}}a_{j}^{a(0)}=x_{i}\p_{t}a_{j}^{(0)}+q_{2}\delta_{ij}a_{t}^{a(0)}
				\end{eqnarray}
				The action \eqref{Electric0} changes under these transformations as
				\begin{eqnarray}
					\delta_{B_{k}}\mathcal{L}^{(0)}=\p_{t}\big(\frac{x_{k}}{2}E_{i}^{a(0)}E_{k}^{a(0)}\big)
				\end{eqnarray}
				So the action is invariant under boost, if the constants $q_{1}$ and $q_{2}$ respectively are 0 and 1.
				
				\subsection*{Magnetic($\delta$=0)}
				{\textbf{Rotation}}:-
				Fields transform as
				\begin{eqnarray}&&
					\delta_{M_{ij}}\xi_{k}^{a}=\big(x_{i}\p_{j}-x_{j}\p_{i}\big)x_{k}^{a}+\delta_{ik}\xi_{j}^{a}-\delta_{jk}\xi_{i}\\&&
					\delta_{M_{ij}}a_{t}^{a(0)}=(x_{i}\p_{j}-x_{j}\p_{i})a_{t}^{a(0)}\\&&
					\delta_{M_{ij}}a_{k}^{a(0)}=(x_{i}\p_{j}-x_{j}\p_{i})a_{k}^{a(0)}+(\delta_{ik}a_{j}^{a(0)}-\delta_{jk}a_{i}^{a(0)})
				\end{eqnarray}
				action \eqref{omag} changes as
				\begin{eqnarray}
					\delta_{M_{ij}}\mathcal{L}^{NLO}=\p_{i}\big(x_{j}\mathcal{L}\big)-\p_{j}\big(x_{i}\mathcal{L}\big)
				\end{eqnarray}
				The action is rotation invariant.\\
				{\textbf{Boost}}:-
				Fields transform as
				\begin{eqnarray}&&
					\delta_{B_{k}}a_{t}^{a}=x_{k}\p_{t}a_{t}^{a}+q_{1}a_{k}^{a}\\&&
					\delta_{B_{k}}a_{i}^{a}=x_{k}\p_{t}a_{i}^{a}+q_{2}\delta_{ik}a_{t}^{a}\\&&
					\delta_{B_{k}}\xi_{i}^{a}=x_{k}\p_{t}\xi_{i}^{a}+q_{3}f_{ik}^{a}
				\end{eqnarray}
				action \eqref{omag} changes as
				\begin{eqnarray}
					\delta_{B_{k}}\mathcal{L}=\p_{t}\big(x_{k}\mathcal{L}\big)
				\end{eqnarray}
				So the action is invariant under boost, if the constant $q_{1}$, $q_{2}$ and $q_{3}$ respectively are 0, -1 and 1.
				\subsection*{Electric($\delta$=1)}
				{\textbf{Rotation}}:-
				Fields transform as
				\begin{eqnarray}&&
					\delta_{M_{ij}}a_{t}^{a(0)}=(x_{i}\p_{j}-x_{j}\p_{i})a_{t}^{a(0)}\\&&
					\delta_{M_{ij}}a_{k}^{a(0)}=(x_{i}\p_{j}-x_{j}\p_{i})a_{k}^{a(0)}+(\delta_{ik}a_{j}^{a(0)}-\delta_{jk}a_{i}^{a(0)})
				\end{eqnarray}
				action \eqref{1electric} change as
				\begin{eqnarray}
					\delta_{M_{ij}}\mathcal{\tilde{L}}^{(0)}=\p_{j}\big(\frac{1}{2}x_{i}\tilde{E}_{k}^{a(0)}\tilde{E}_{k}^{a(0)}\big)-\p_{i}\big(\frac{1}{2}x_{j}\tilde{E}_{k}^{a(0)}\tilde{E}_{k}^{a(0)}\big)
				\end{eqnarray}
				So the action is invariant under rotation.
				\\
				{\textbf{Boost}}
				Fields transform as 
				\begin{eqnarray}&&
					\delta_{B_{i}}a_{t}^{a(0)}=x_{i}\p_{t}a_{t}^{a(0)}\\&&
					\delta_{B_{i}}a_{j}^{a(0)}=x_{i}\p_{t}a_{j}^{(0)}+\delta_{ij}a_{t}^{a(0)}
				\end{eqnarray}
				action \eqref{1electric} change as
				\begin{eqnarray}
					\delta_{B_{i}}\mathcal{\tilde{L}}^{(0)}=\p_{t}\big(\frac{1}{2}x_{i}\tilde{E}_{k}^{a(0)}\tilde{E}_{k}^{a(0)}\big)
				\end{eqnarray}
				
				So the action is invariant under boost. 
				\subsection*{Magnetic($\delta$=1)}
				\textbf{{Rotation}}:-
				Fields transform as
				\begin{eqnarray}&&
					\delta_{M_{ij}}\xi_{k}^{a}=\big(x_{i}\p_{j}-x_{j}\p_{i}\big)x_{k}^{a}+\delta_{ik}\xi_{j}^{a}-\delta_{jk}\xi_{i}\\&&
					\delta_{M_{ij}}a_{t}^{a(0)}=(x_{i}\p_{j}-x_{j}\p_{i})a_{t}^{a(0)}\\&&
					\delta_{M_{ij}}a_{k}^{a(0)}=(x_{i}\p_{j}-x_{j}\p_{i})a_{k}^{a(0)}+(\delta_{ik}a_{j}^{a(0)}-\delta_{jk}a_{i}^{a(0)})
				\end{eqnarray}
				action\eqref{1magaction} changes as
				\begin{eqnarray}
					\delta_{M_{ij}}\mathcal{L}^{NLO}=\p_{i}\big(x_{j}\mathcal{L}\big)-\p_{j}\big(x_{i}\mathcal{L}\big)
				\end{eqnarray}
				So the action is invariant under rotation.\\
				{\textbf{Boost}}:-
				Fields transform as
				\begin{eqnarray}&&
					\delta_{B_{k}}a_{t}^{a}=x_{k}\p_{t}a_{t}^{a}\\&&
					\delta_{B_{k}}a_{i}^{a}=x_{k}\p_{t}a_{i}^{a}-\delta_{ik}a_{t}^{a}\\&&
					\delta_{B_{k}}\xi_{i}^{a}=x_{k}\p_{t}\xi_{i}^{a}+\tilde{f}_{ik}^{a}
				\end{eqnarray}
				Action\eqref{1magaction} change as
				\begin{eqnarray}
					\delta_{B_{k}}\mathcal{L}=\p_{t}\big(x_{k}\mathcal{L}\big)
				\end{eqnarray}
				So the action is invariant under boost. 
				
				\section{Charge Algebra}\label{Charge Algebra app}
				In sec.\ref{Noether charge and algebra} have Noether's charge and discussion on charge algebra. In this appendix, we give a charge in pre-symplectic language with some examples.
				\subsection*{Electric Sector}
				 Using the expression of $\Theta$, we can define the Poisson bracket  for the electric sector as
				\begin{eqnarray}\label{Poisson bracket}
					\Omega(\delta_{1},\delta_{2})=\int d^{3}x\big[\delta_{1}\Theta(\delta_{2})-\delta_{2}\Theta(\delta_{1})\big]=\int  d^{3}x\big[\delta_{1}a_{i}^{a(0)}\delta_{2}E_{i}^{a(0)}-\delta_{2}a_{i}^{a(0)}\delta_{1}E_{i}^{a(0)}\big]
				\end{eqnarray} 
				Let see some commutation of Conformal Carrollian algebra.
			\paragraph*{$[P,P]$:}
			\begin{eqnarray}
				\delta_{P_{i}}a_{j}^{a(0)}=\p_{i}a_{j}^{a(0)},\quad 	\delta_{P_{i}}a_{t}^{a(0)}=\p_{i}a_{t}^{a(0)}
			\end{eqnarray}
			Using these expression in the Eq.\eqref{Poisson bracket} we will get
			\begin{eqnarray}&&
            \hspace*{-1.8cm} \Omega(\delta_{P_{l}},\delta_{P_{k}})=\int  d^{3}x\big[\p_{l}a_{i}^{a(0)}\p_{k}E_{i}^{a(0)}-\p_{k}a_{i}^{a(0)}\p_{l}E_{i}^{a(0)}\big]\non\\&&
             =\int  d^{3}x\big[\p_{l}(a_{i}^{a(0)}\p_{k}E_{i}^{a(0)})-\p_{k}(a_{i}^{a(0)}\p_{l}E_{i}^{a(0)})\big]=0
			\end{eqnarray}
		Last equality is zero because of the total derivative in the previous step.
		\paragraph*{$[P,M]$:
		\bes
		\begin{eqnarray}\label{emfalgebra}&&
		\delta_{M_{f}}a_{t}^{a(0)}=f(x)\p_{t}a_{t}^{a(0)},\quad
		\delta_{M_{f}}a_{i}^{a(0)}=f(x)\p_{t}a_{i}^{a(0)}+a_{t}^{a(0)}\p_{i}f(x)
		\end{eqnarray}
		\ees}
			\begin{eqnarray}&&
		\hspace*{-1.8cm} \Omega(\delta_{P_{l}},\delta_{M_{f(x)}})=\int  d^{3}x\big[\p_{l}a_{i}^{a(0)}f(x)\p_{t}E_{i}^{a(0)}-(f(x)\p_{t}a_{i}^{a(0)}+a^{a(0)}_{t}\p_{l}f(x))\p_{l}E_{i}^{a(0)}\big]\non\\&&
		=\int  d^{3}x\big[\p_{l}f(x)E_{i}^{a(0)}E_{i}^{a(0)}+f(x)\p_{l}a_{i}^{a(0)}\p_{t}E_{i}^{a(0)}+f(x)a_{t}^{a(0)}\p_{t}\p_{i}E_{i}^{a(0)}\big]\non\\&&
		=Q_{electric}(M_{h}) \quad \text{where} \quad h=\p_{l}f(x)
		\end{eqnarray}
		\paragraph*{[$D,M$]:}

		\begin{eqnarray}&&
		\hspace*{-1.8cm} \Omega(\delta_{D},\delta_{M_{f(x)}})=Q_{electric}(M_{h_{1}}) \quad \text{where} \quad h_{1}= -f(x)+x^{k}\p_{k}f(x)
		\end{eqnarray}
		
	\paragraph*{$[K_{i},M_{f}]$}
	\begin{eqnarray}
		\Omega(\delta_{K_{i}},\delta_{M_{f}})=Q_{electric}(M_{h_{2}}),\quad \text{where}\quad h_{2}=(2x_{i}x^{k}\p_{k}-x^{k}x_{k}\p_{i}-2x_{i})f(x)
	\end{eqnarray}
		Expression of the $Q_{electric}$ is given in  Sec.\ref{Noether charge and algebra}. 
		\section*{Magnetic Sector}
		Using the expression of $\Theta$, we can define the Poisson bracket  for the magnetic sector as
		\begin{eqnarray}\label{Poisson bracket Magnetic}
			\Omega(\delta_{1},\delta_{2})=\int d^{3}x\big[\delta_{1}\Theta(\delta_{2})-\delta_{2}\Theta(\delta_{1})\big]=\int  d^{3}x\big[\delta_{1}a_{i}^{a}\delta_{2}\xi_{i}^{a}-\delta_{2}a_{i}^{a}\delta_{1}\xi_{i}^{a}\big]
		\end{eqnarray} 
	
	Similar to electric case we can see
	\paragraph*{$[P,P]$:}
	Using these expression in the Eq.\eqref{Poisson bracket} we will get
	\begin{eqnarray}&&
		\hspace*{-1.8cm} \Omega(\delta_{P_{l}},\delta_{P_{k}})=0
	\end{eqnarray}
	Last equality is zero because of the total derivative in the previous step.
	\paragraph*{$[P,M]$}:
		 Using the transformation given in previous section
		
	\begin{eqnarray}&&
		\hspace*{-1.8cm} \Omega(\delta_{P_{l}},\delta_{M_{f(x)}})=Q_{Magnetic}(M_{h})\quad \text{where} \quad h=\p_{l}f(x)
	\end{eqnarray}
	\paragraph*{[$D,M$]:}
	
	\begin{eqnarray}&&
		\hspace*{-1.8cm} \Omega(\delta_{D},\delta_{M_{f(x)}})=Q_{Magnetic}(M_{h})\quad \text{where} \quad h_{1}= -f(x)+x^{k}\p_{k}f(x)
	\end{eqnarray}
	
	\paragraph*{$[K_{i},M_{f}]$}
	\begin{eqnarray}
		\Omega(\delta_{K_{i}},\delta_{M_{f}})=Q_{Magnetic}(M_{h_{1}}),\quad \text{where}\quad h_{2}=(2x_{i}x^{k}\p_{k}-x^{k}x_{k}\p_{i}-2x_{i})f(x)
	\end{eqnarray}
Expression of the $Q_{magnetic}$ is given in  Sec.\ref{Noether charge and algebra}.  

	\section{Discussion on previous work on Carrollian Yang-Mills theory}
In \cite{Bagchi:2016bcd}, authors discussed Carrollian Yang-Mills theory at the level of equations of motion. In their analysis for the $SU(2)$ theory, there are four different sectors of Carrollian Yang-Mills equations of motion. For the details discussion readers are encouraged to see the references mentioned above. Here we will do a similar analysis and see how we can relate our results to the previous analysis. The relativistic equations of motion is
\begin{eqnarray}
	\p_{{\mu}}F_{\mu\nu}^{a}+gf^{abc}A_{\mu}^{b}F_{\mu\nu}^{c}=D_{\mu}F_{\mu\nu}^{a}=0,
\end{eqnarray}
we can write temporal and spatial part as
\begin{eqnarray}&&
	\p_{{i}}F_{i 0}^{a}+gf^{abc}A_{i}^{b}F_{i0}^{c}=D_{i}F_{i0}^{a}=0\\&&
	\p_{{i}}F_{ij}^{a}+gf^{abc}A_{i}^{b}F_{ij}^{c}=D_{i}F_{ij}^{a}=0.
\end{eqnarray}
To derive the Carrollian Yang-Mills equations of motion using the formalism discussed in  \cite{Bagchi:2016bcd} we have to scale $t,x$ and  all the fields of the theory along with coupling ($g$) as
\begin{eqnarray}
	x_{i}\rightarrow\epsilon^{\beta}x_{i},x_{0}\rightarrow\epsilon^{\beta+1}t,\, A_{i}^{a}\rightarrow \epsilon^{\alpha+1}a_{i}^{a},\,A_{0}^{a}\rightarrow \epsilon^{\alpha}a_{t}^{a},\,g\rightarrow\epsilon^{\gamma}g,\, \text{with} \,\,\epsilon\rightarrow 0
\end{eqnarray}
In this limit the consistent equations of motions are
\begin{eqnarray}&&
	\p_{{i}}E_{i}^{a}+gf^{abc}a_{i}^{b}E_{i}^{c}=D_{i}E_{i}^{a}=0\\&&
	\p_{{i}}f_{ij}^{a}+gf^{abc}a_{i}^{b}f_{ij}^{c}=D_{i}f_{ij}^{a}=0\\&&
	\hspace*{2cm}\text{if} \quad \gamma=-(\alpha+\beta+1)\non.
\end{eqnarray} 
where
$E_{i}^{a}=\p_{t}a_{i}^{a}-\p_{i}a_{t}^{a}+gf^{abc}a_{t}^{a}a_{i}^{a}$, $f_{ij}^{a}=\p_{i}a_{j}^{a}-\p_{j}a_{i}^{a}+gf^{abc}a_{i}^{a}a_{j}^{a}$. These equations of motion are same as eq.\eqref{EOM}.
The equations of motion we get here by scaling of fields are reproduced from the electric sector action discussed in the section.\ref{Non-trivial electric}. 

In sec.\ref{trivial electric} we have another electric sector of Carrollian Yang-Mills theory which is copies of the electric sector of Carrollian abelian theory. The equations of motion of this electric sector are computed previously in \cite{Bagchi:2016bcd}.

This paper's magnetic sector equations of motion do not match the previous works done in \cite{Bagchi:2016bcd} because these results derive from the relativistic theory with a Lagrange Multiplier.


				\bibliographystyle{JHEP}
				\bibliography{CYM}
			\end{document}